\documentclass[prd,twocolumn,showpacs,preprintnumbers,nofootinbib,superscriptaddress]{revtex4}  
\usepackage{amsmath}
\usepackage{rsfs}
\usepackage{bm}
\usepackage[pdftex]{graphicx}
\usepackage{color}

\usepackage{ulem}
\usepackage{cleveref}

\allowdisplaybreaks

\renewcommand{\theequation}{{\arabic{equation}}}

\newcommand{\beq}{\begin{eqnarray*}}
\newcommand{\eeq}{\end{eqnarray*}}

\begin{document}
\preprint{CHIBA-EP-244, KEK Preprint 2019-48}

\title{Double-winding Wilson loops
in $SU(N)$ lattice Yang-Mills gauge theory
}

\author{Seikou Kato}
\email{skato@oyama-ct.ac.jp}
\affiliation{Oyama National College of Technology, Oyama 323-0806, Japan}

\author{Akihiro Shibata}
\email{akihiro.shibata@kek.jp}
\affiliation{Computing Research Center, High Energy Accelerator Research Organization (KEK), Oho 1-1, 
Tsukuba 305-0801, Japan}

\author{Kei-Ichi Kondo}
\email{kondok@faculty.chiba-u.jp}
\affiliation{Department of Physics,  
Graduate School of Science, 
Chiba University, Chiba 263-8522, Japan
}

\begin{abstract}

We study  double-winding Wilson loops in $SU(N)$ lattice Yang-Mills gauge theory by using both strong coupling expansions and numerical simulations. 
First, we examine how the area law falloff of a ``coplanar'' double-winding Wilson loop average 
depends on the number of color $N$.
Indeed, we find that a coplanar double-winding Wilson loop average obeys a novel ``max-of-areas law'' for $N=3$ and 
the sum-of-areas law for $N\geq 4$, although we reconfirm the difference-of-areas law for $N=2$.
Second, we examine a ``shifted''  double-winding Wilson loop, where the two constituent loops are displaced from one another in a transverse direction.
We evaluate its average by changing the distance of a transverse direction and we find that the long distance behavior does not depend on  the number of color $N$, while the short distance behavior depends strongly on $N$.

\end{abstract}


\maketitle
 



\section{Introduction}

What is the true mechanism for \textit{quark confinement} is not yet confirmed and still under the debate, although more than 50 years have passed since \textit{quark model} was proposed by Gell-Mann  \cite{Gell-Mann} in the beginning of 1960s. 
In the 1970s, however, the \textit{dual superconductor picture}  was already proposed by Nambu, 't Hooft and Mandelstam  \cite{dualsuper} as a mechanism for quark confinement. 
In fact, validity of the dual superconductor picture was confirmed for $U(1)$ pure gauge theory \cite{Polyakov75}, Georgi-Glashow model \cite{Polyakov77} and $\mathcal{N}=2$ supersymmetric Yang-Mills theory \cite{SW94}, although it is not yet confirmed for the ordinary non-supersymmetric Yang-Mills theory \cite{YM54} and quantum chromodynamics (QCD). 
Therefore, the dual superconductor picture is now regarded as one of the most promising scenarios for quark confinement, although this does not deny the existence of the other mechanics for quark confinement. 
See e.g., \cite{Bali01,Greensite03,KKSS15} for reviews. 

In order to establish the dual superconductor scenario, the most difficult issue to be resolved first of all is to guarantee the existence of \textit{magnetic monopoles} in the pure non-Abelian Yang-Mills gauge theory, which is different from the 't Hooft--Polyakov magnetic monopole \cite{tHP74} in the gauge-scalar model. 
This issue was circumvented by using the method called the \textit{Abelian projection} proposed by 't Hooft \cite{tHooft81}.  The Abelian projection is a gauge fixing which explicitly breaks the original gauge group into its maximal torus subgroup where  color symmetry is also broken.  By the Abelian projection, magnetic monopoles of the Abelian type \cite{Dirac31,WY75} are indeed realized, but the resulting theory is distinct from the original gauge theory with the non-Abelian gauge group. 
To avoid the gauge artifact, we must find a procedure which enables one to define magnetic monopoles in a gauge-invariant way. 
This issue was solved recently for the Yang-Mills theory with the gauge group $SU(N)$ and any semi-simple compact gauge group \cite{MK05}, by using the non-Abelian Stokes theorem for the Wilson loop operator and the new reformulation of the Yang-Mills theory based on the new field variables obtained by  change of variables through the gauge covariant field decomposition of the Cho-Duan-Ge-Faddeev-Niemi-Shabanov \cite{DG79,Cho80,FN98,Shabanov99,Cho80c,FN99a,KMS05,KMS06}. 
See \cite{KKSS15} for a recent review. 

\begin{figure}[tbp]
\begin{center}
\includegraphics[height=2.5cm]{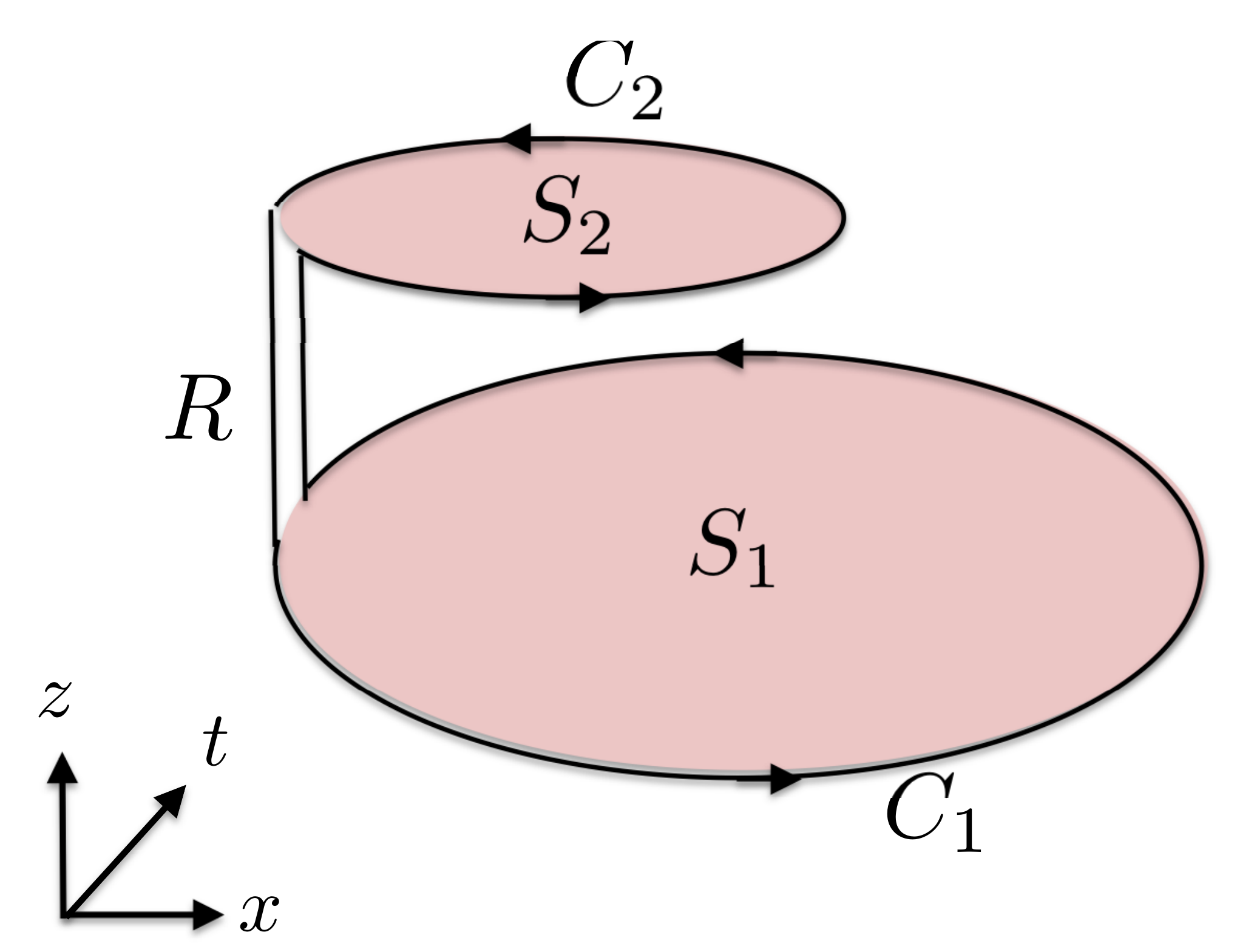}
\includegraphics[height=2.5cm]{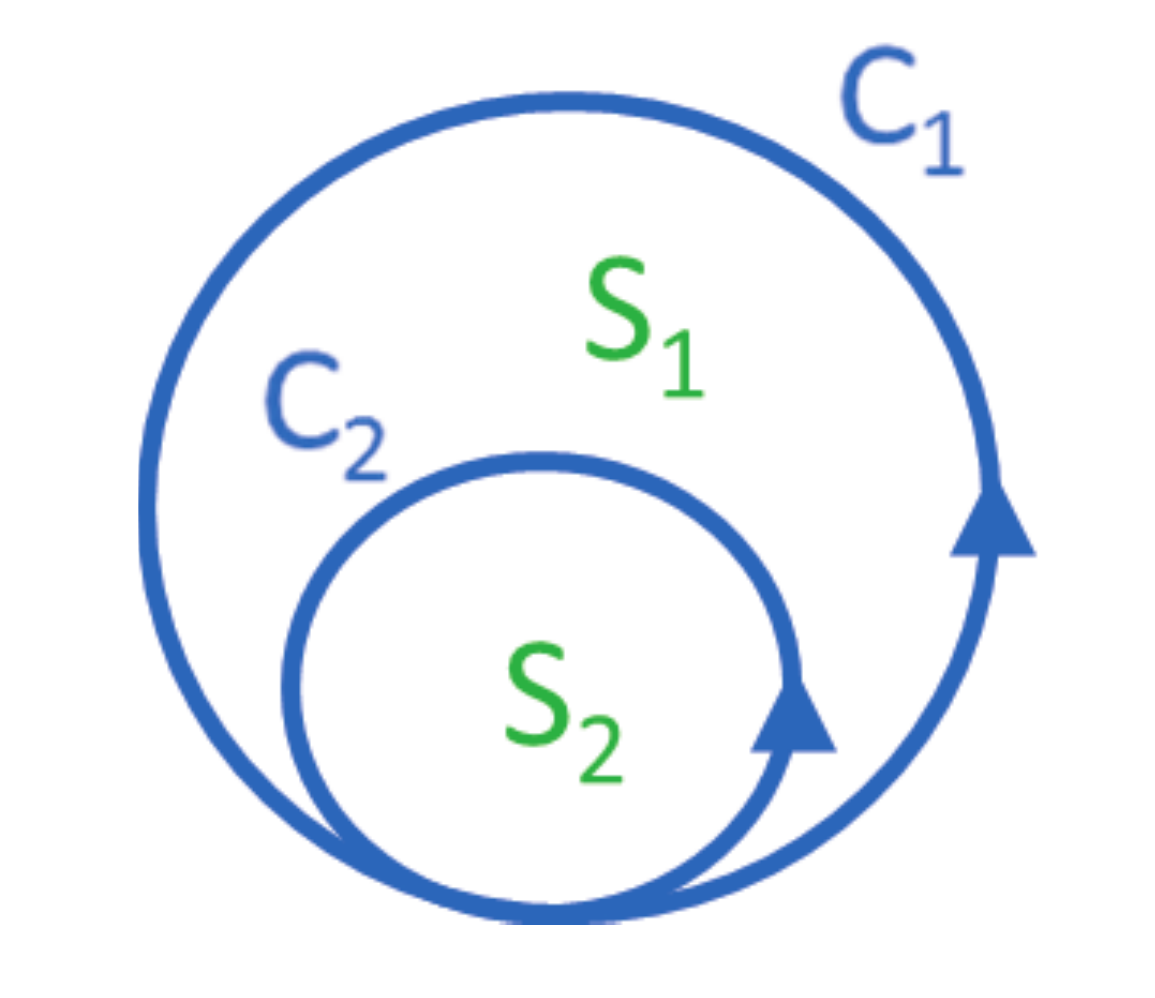}
\caption{  
The double-winding Wilson loops. (left) A ``shifted'' double-winding Wilson loop $W( C=C_1\times C_2 )$ composed of the two loops $C_1$ and $C_2$ which lie in planes parallel to the $x-t$ plane, 
but are displaced from one another in the $z$-direction by distance $R$.
(right) a ``coplanar'' double-winding Wilson loop $W( C=C_1\times C_2 )$ as the limit $R=0$ of the ``shifted'' double-winding Wilson loop. 
}
\label{Dw-fig1}
\end{center}
\end{figure}

However, these achievements do not necessarily means that the dual superconductivity is the unique scenario for understanding quark confinement. 
Recently, Greensite and H\"ollwieser  \cite{GH15} introduced a ``double-winding'' Wilson loop operator 
in lattice gauge theory \cite{Wilson74} to examine possible mechanisms for quark confinement.
The \textit{double-winding} Wilson loop operator $W(C=C_1 \times C_2)$ is a path-ordered product of (gauge) link  variables $U_\ell \in SU(N)$ along a 
closed contour $C$ which is composed of two loops $C_1$ and $C_2$, 
\begin{align}
W( C ) \equiv {\rm tr} [\prod_{\ell \in C}U_\ell], \quad  C=C_1 \times C_2.
\label{chap3-1}
\end{align}
See Fig.\ref{Dw-fig1}.
A more general ``shifted'' double-winding loop is introduced in such a way that the two loops $C_1$ and $C_2$ lie in planes parallel to the $x-t$ plane, 
but are displaced from one another in the transverse direction, e.g., $z$ by distance $R$, and are connected by lines running parallel to the $z$-axis.
In the non-shifted case $R=0$, the two loops $C_1$ and $C_2$ lie in the same plane, which we call \textit{coplanar}.
We denote by $S_1$ and $S_2$ the minimal areas bounded by loops $C_1$ and $C_2$, respectively.
Note that the double-winding Wilson loop operator is defined in a gauge invariant manner, irrespective of shifted $R \not= 0$ or coplanar $R=0$.


\begin{figure}[tbp]
\begin{center}
\includegraphics[height=5.0cm]{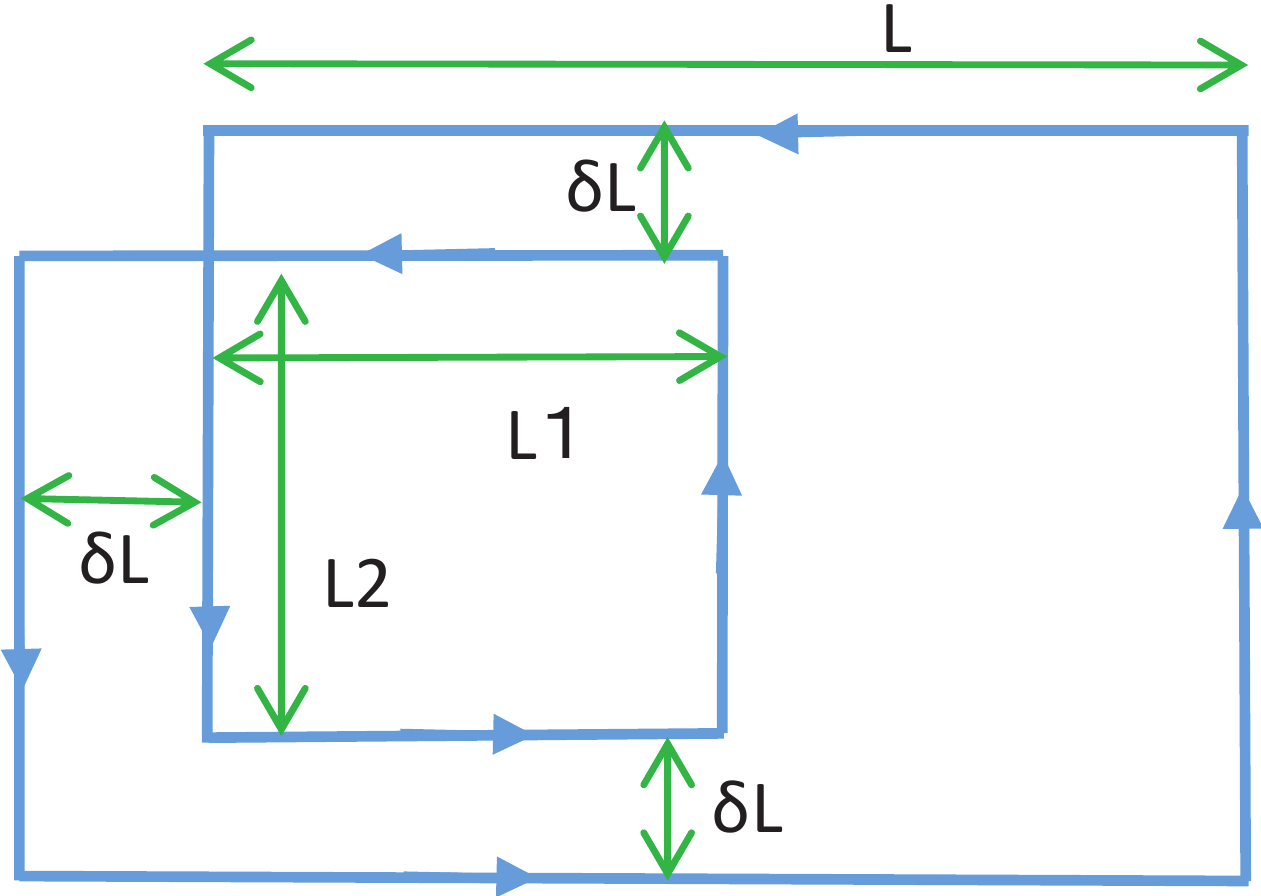}
\caption{ 
The setting up of a coplanar double-winding Wilson loop.
}
\label{Dw-fig3}
\end{center}
\end{figure}

In \cite{GH15}, they investigated the area   ($S_1$ and $S_2$) dependence of the expectation value $\left< W(C=C_1 \times C_2) \right>$ of a double-winding Wilson loop operator $W(C=C_1 \times C_2)$ for the $SU(2)$ gauge group.
Consequently, it has been shown in a numerical way that both the original $SU(2)$ lattice gauge theory and   
center vortex model obey the \textit{difference-of-areas} $(S_1-S_2)$ law, while the Abelian-projected model obeys the \textit{sum-of-areas} $(S_1+S_2)$ law. 
In the coplanar case $R=0$, a double-winding loop has been set up as given in Fig.\ref{Dw-fig3}.
In order to discriminate difference-of-areas and sum-of-areas laws, it is efficient to measure the $L_1$-dependence of a coplanar double-winding Wilson loop average  $\langle W( C=C_1\times C_2 ) \rangle$, with the other lengths $L$, $L_2$, and $\delta L$ being fixed.
For simplicity, we set  $\delta L=0$. 
Then $S_1(= L \times L_2)$ and $S_2(= L_1\times L_2)$ are the minimal areas of rectangular loops $C_1$ and $C_2$, respectively.
We assume $S_1\geq S_2$ for definiteness hereafter.
 If  $\langle W( C_1\times C_2 ) \rangle$ obeys the difference-of-areas law:
\begin{align}
\langle W( C_1\times C_2 ) \rangle
\simeq & \exp[-\sigma |S_1-S_2|]
\nonumber\\
=& \exp[-\sigma L_2(L-L_1)],
\label{chap4-1}
\end{align}
then $\ln \langle W(  C_1\times C_2 ) \rangle$ must linearly increase  in $L_1$ as $L_1$ increases.
On the other hand, if $\langle W( C_1\times C_2 ) \rangle$ obeys the sum-of-areas law:
\begin{align}
\langle W( C_1\times C_2 ) \rangle
 \simeq & \exp[-\sigma' (S_1+S_2)]
\nonumber\\
=& \exp[-\sigma' L_2(L+L_1)],
\label{chap4-2}
\end{align}
then $\ln \langle W(  C_1\times C_2 ) \rangle$ must linearly decrease  in $L_1$ as $L_1$ increases.

\begin{figure}[tbp]
\begin{center}
\includegraphics[height=5.2cm]{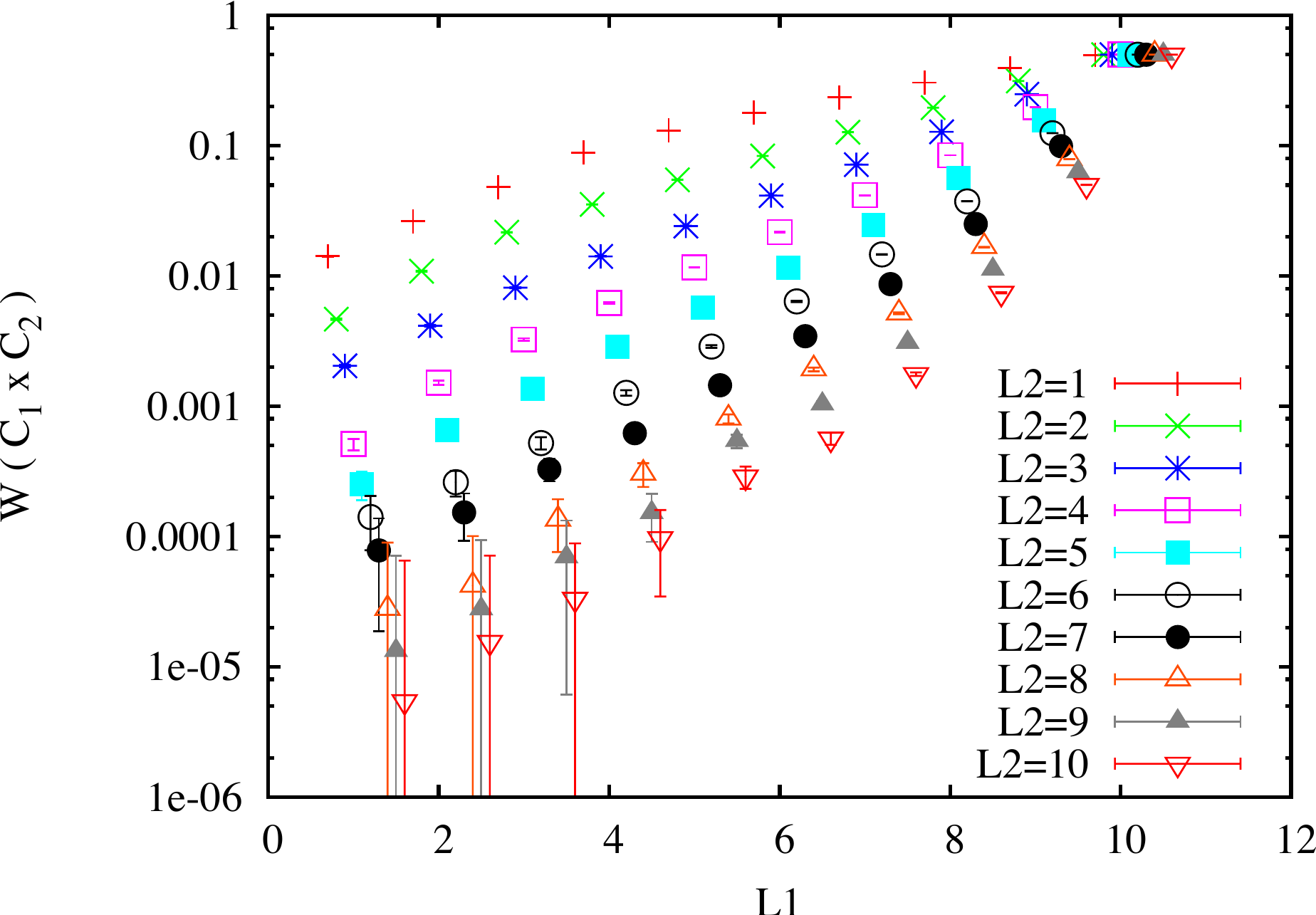}
\includegraphics[height=5.2cm]{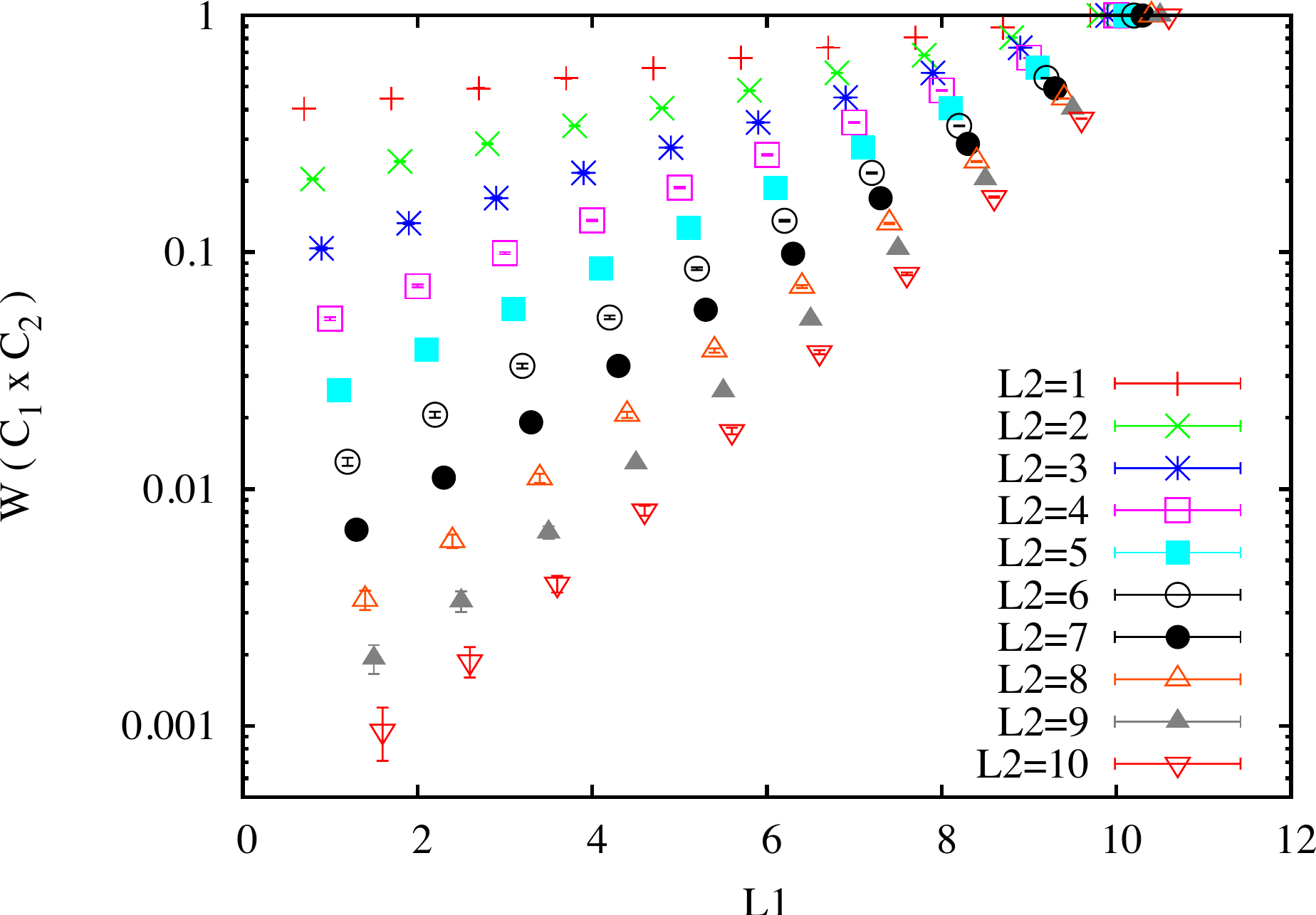}
\includegraphics[height=5.2cm]{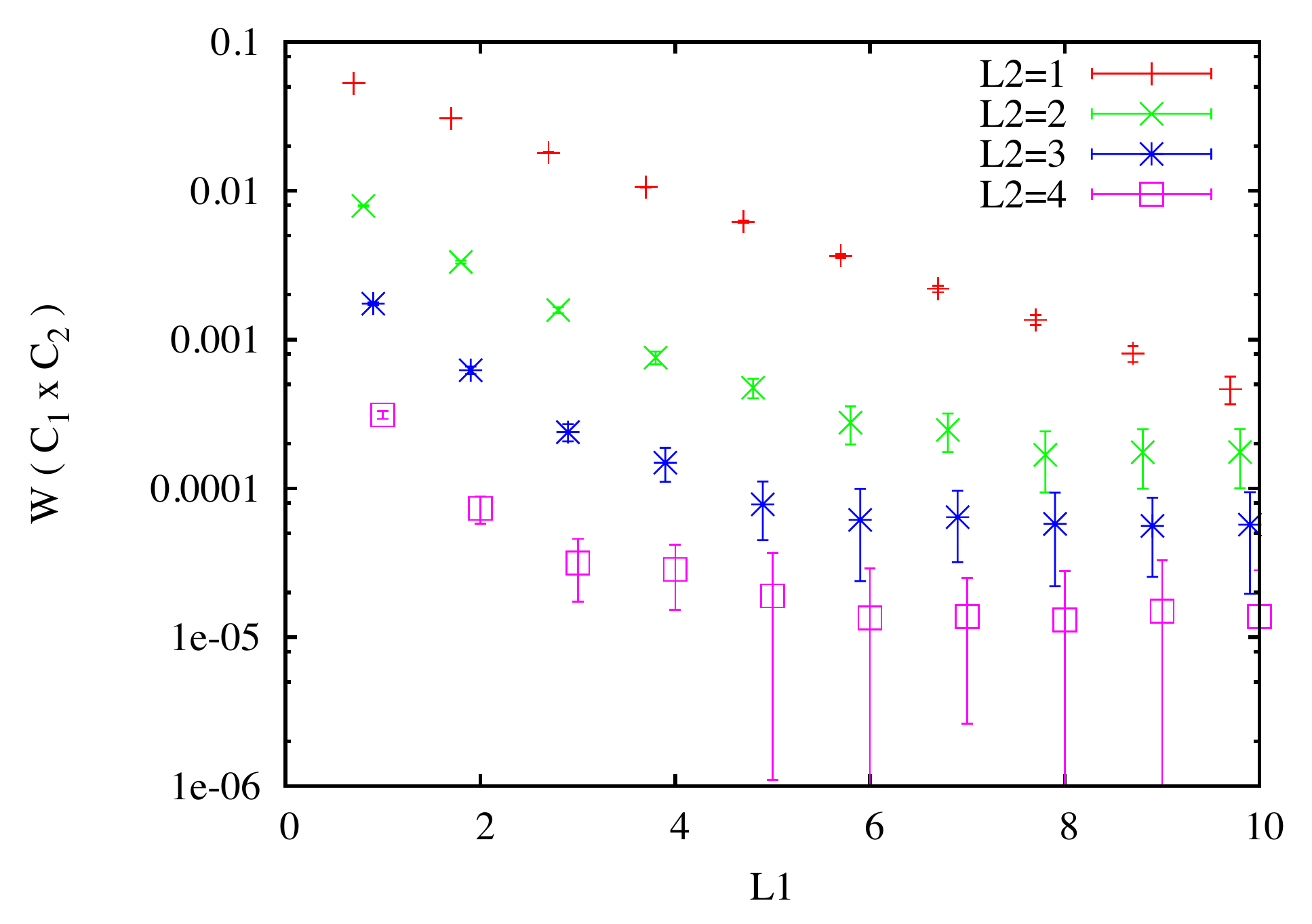}
\caption{  
$L_1$ dependence of a coplanar double-winding Wilson loop average  $\langle W( C=C_1\times C_2 ) \rangle$   (top panel) for the original $SU(2)$ field, [reproduced from  Fig.7.(a) in \cite{GH15}],  
(middle panel)  for center vortex [reproduced from  Fig.7.(c) in \cite{GH15}], 
(bottom panel)  for Abelian degree of freedom, [reproduced  from  Fig.8.(c) in \cite{GH15}]. 
}
\label{Dw-fig4}
\end{center}
\end{figure}

The numerical evidences were obtained as given in 
Fig.\ref{Dw-fig4} which summarizes their results for $L_1$ dependence of $\ln \langle W(  C_1\times C_2 ) \rangle$ with the other lengths being fixed, e.g., $L=10$, $L_2=1$, $\delta L=0$, based on  numerical simulations performed on a lattice of size $20^4$ at $\beta=2.4$.  
These results certainly show both the original $SU(2)$ gauge field and center vortex lead to the difference-of-areas law, while Abelian-projected configurations lead to the sum-of-areas law.


\begin{figure}[tbp]
\begin{center}
\includegraphics[height=1.5cm]{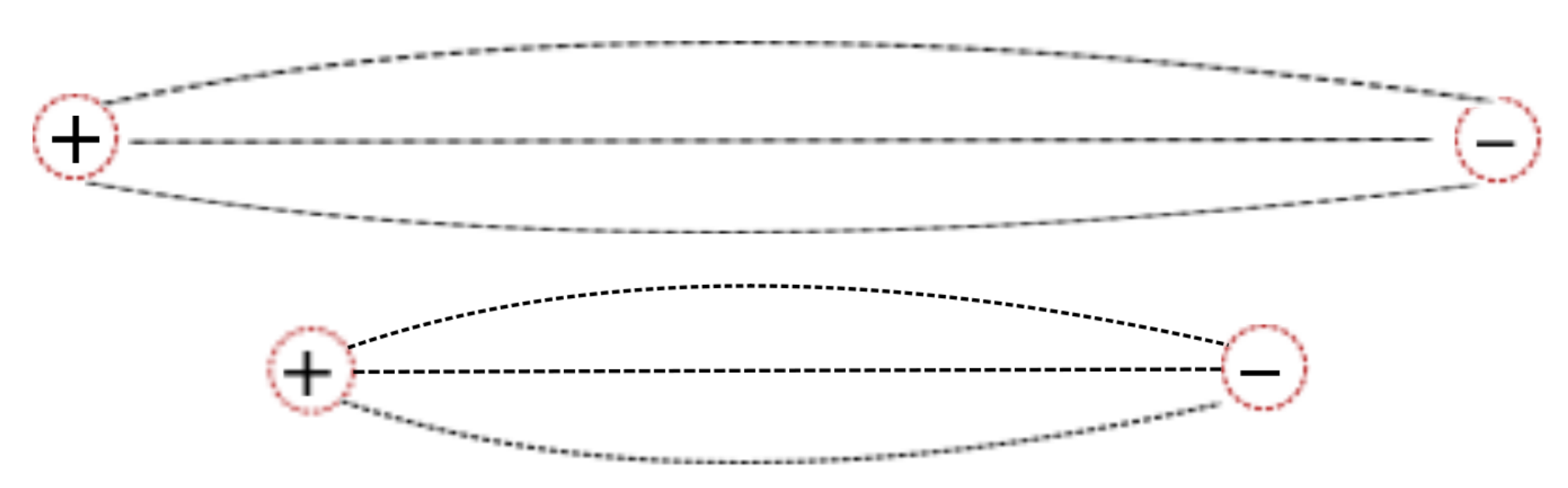}

\vspace{1.0cm}

\includegraphics[height=1.2cm]{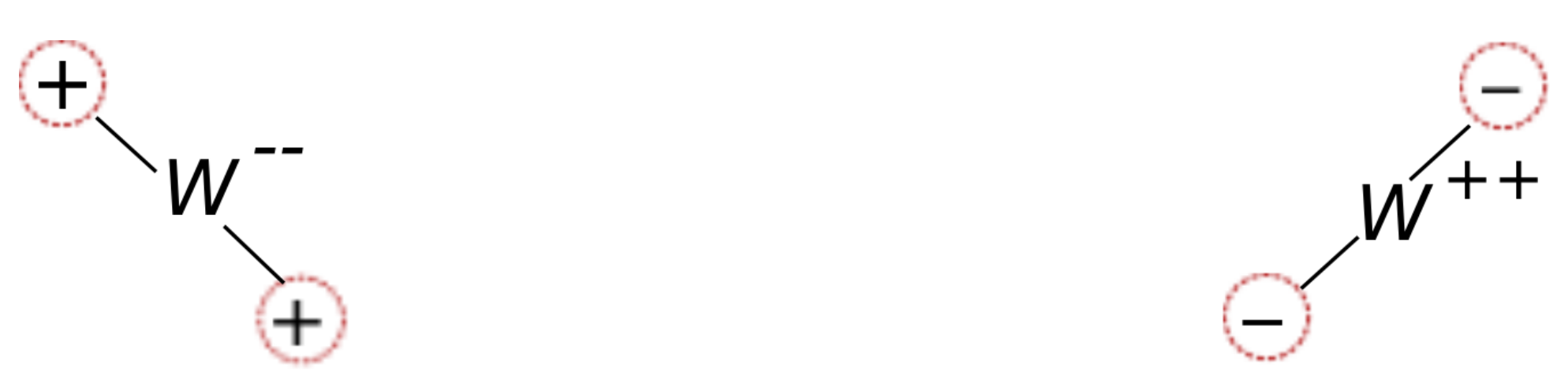}
\caption{(top panel) Interactions between flux tubes generated by two pairs of a quark and an antiquark, leading to the sum-of-areas law [reproduced from Fig.3 in \cite{GH15}]. (bottom panel) W boson neutralizes the widely separated positive and negative charges, leading to the difference-of-areas law in $SU(2)$, [reproduced from Fig.11 in \cite{GH15}].}
\label{Dw-fig2}
\end{center}
\end{figure}

From a physical point of view, a double-winding Wilson loop can be interpreted as a probe for studying interactions between two pairs of a particle and an antiparticle. Then differences among three cases are understood as follows. 
In the Abelian model, a particle and an antiparticle in a pair are respectively connected by the electric flux with the length of $L$ and $L_1$, as indicated in the top panel of Fig.\ref{Dw-fig2}.
The total energy of flux tubes shifted by $R>0$ becomes $\sigma' (L+L_1)$, where $\sigma' $ is a string tension, if the flux-flux interactions are neglected. This argument will give a reason why the Abelian model gives the sum-of-areas law.
Moreover, they argue that even in the limit $R \to 0$   the sum-of-areas law remains unchanged in the  Abelian model, because electric flux tubes tend to repel each other and they can not coincide in the  type II dual superconductor.

For the $SU(2)$ gauge theory, they argue that the ``$W$ bosons'' play the crucial role, since they are off-diagonal components of the $SU(2)$ gauge field which are not included in the Abelian model. $W$ bosons have charged components $W^{--}$ and $W^{++}$ with respect to the Abelian $U(1)$ group.  They explain that charged off-diagonal components $W^{--}$ and $W^{++}$ of the $SU(2)$ gauge field neutralize respectively positive and negative static charges. Consequently, flux tubes exist only for connecting two positive charges and two negative static charges, which leads to difference-of-areas law.
See the bottom panel of Fig.\ref{Dw-fig2}.

In the vortex picture, if a vortex pierces the minimal area of a loop, it will multiply the holonomy around the loop by a factor $-1$.  Therefore, if a vortex pierces two loops $C_1$ and $C_2$ simultaneously, it gives a trivial effect. The non-trivial result is obtained only if a vortex pieces the  non-overlapping region $S_1-S_2$. This leads to difference-of-areas law.

Quite recently, Matsudo and Kondo \cite{matsudo-kondo} have investigated a  double-winding, a triple-winding, and general multiple-winding Wilson loops in the continuum $SU(N)$ Yang-Mills theory. 
They have found that a coplanar double-winding $SU(3)$ Wilson loop average follows a novel area law which is neither difference-of-areas law nor sum-of-areas law, and that sum-of-areas law is allowed for $SU(N)$ ($N\geq 4$), if the string tension is assumed to obey the Casimir scaling for quarks in the higher representations.

In this way, the study of double-winding Wilson loops itself is interesting because it can be used to test the  confinement mechanism in QCD. 
Moreover, it is worth considering the interactions between two  color flux tubes.
In this paper, we investigate  both ``coplanar''  and  ``shifted''  double-winding Wilson loops in $SU(N)$ lattice Yang-Mills gauge theory by using both strong coupling expansion and numerical simulations.

In this paper, 
we show that 
the ``coplanar'' double-winding Wilson loop average has the $N$ dependent area law falloff:  
``max-of-areas law'' for $N=3$ and sum-of-areas law for $N\geq 4$, 
which add a new result to the known difference-of-areas law for an $N=2$ ``coplanar'' double-winding Wilson loop average.
Moreover, we investigate the behavior of a ``shifted''  double-winding Wilson loop average as a function of the distance in a transverse direction and find that the long distance behavior does not depend on the number of color $N$, while the short distance behavior depends on $N$.

This article is organized as follows.
In section II, we examine how the area law falloff of a ``coplanar'' double-winding Wilson loop average 
depends on the number of color $N$.
In section III, we examine a ``shifted''  double-winding Wilson loop, where the two constituent loops are displaced from one another in a transverse direction, especially evaluate its average by changing the distance of a transverse direction.
The final section IV is devoted to conclusion and discussion.   
We also discuss the validity of the Abelian operator studied in  \cite{GH15}.
Recently, there are numerical evidences that the dual superconductor for $SU(2)$ and $SU(3)$ lattice Yang-Mills theory is type I
\cite{kato2015}, although they explain sum-of-areas law on the basis of type II superconductor.  
We should study the interaction between two flux tubes in the limit $R \to 0$, in case of type I superconductor. 

\section{A ``coplanar'' double-winding Wilson loop}

First of all, we consider the coplanar case $R=0$ of a double-winding Wilson loop  in the $SU(N)$ lattice Yang-Mills 
gauge theory, as indicated in Fig.\ref{Dw-fig3}.
For simplicity, we set  $\delta L=0$. 
Let $S_1(= L \times L_2)$ and $S_2(= L_1\times L_2)$ be the minimal areas of rectangular loops $C_1$ and $C_2$, respectively.
We assume $S_1\geq S_2$ for definiteness hereafter.

\subsection{strong coupling expansion}

\begin{figure}[tbp]
\begin{center}
\includegraphics[height=3.0cm]{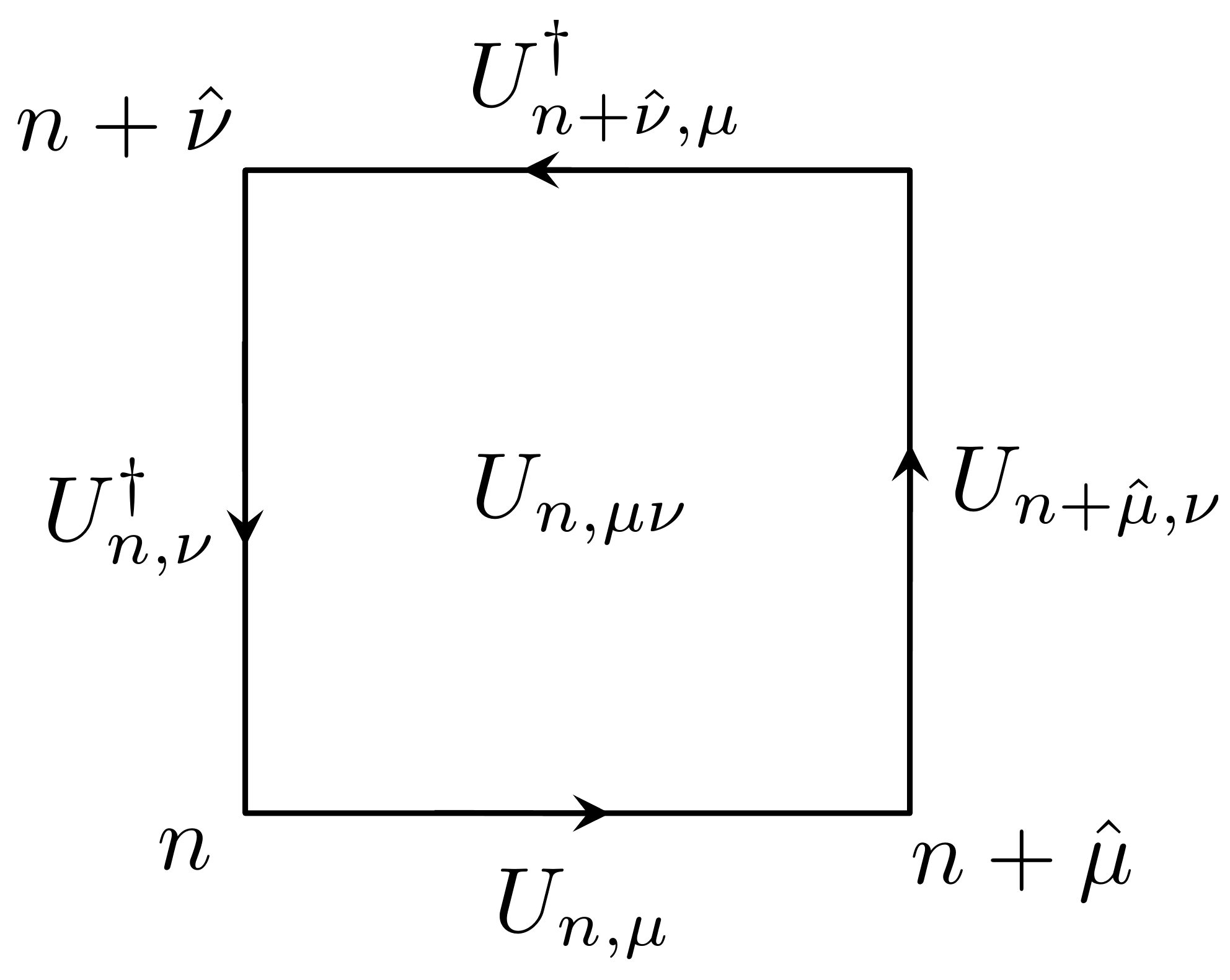}
\includegraphics[height=2.5cm]{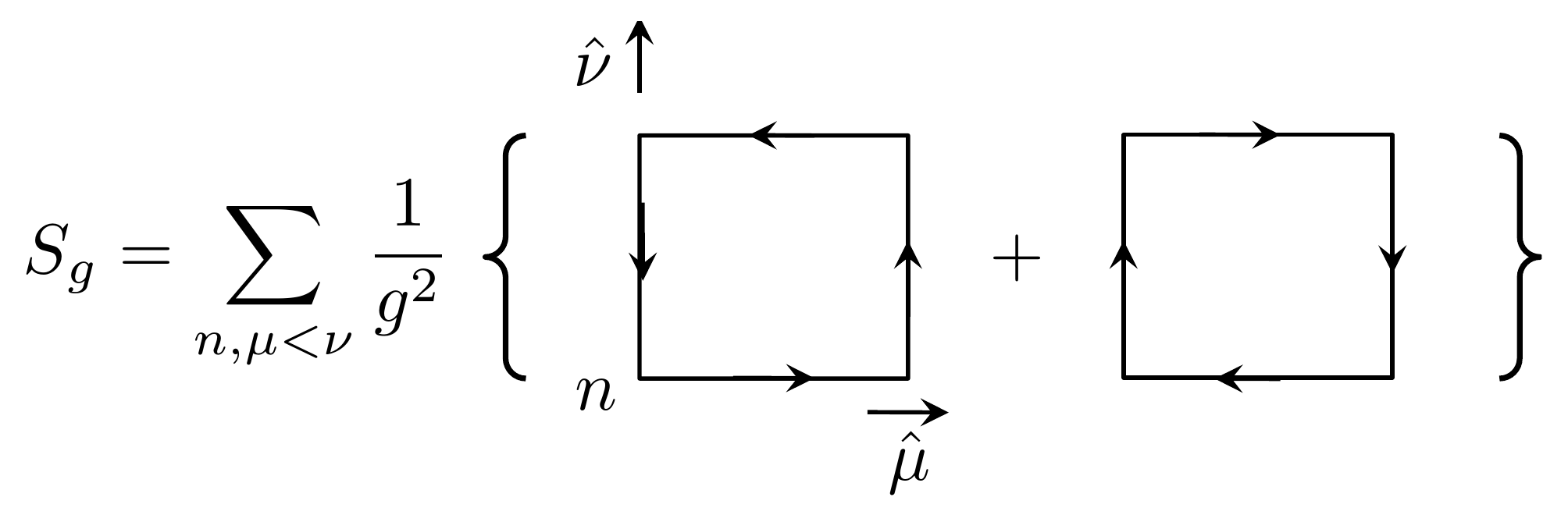}
\caption{ 
(top panel) a plaquette variable $U_{n,\mu\nu}$, 
(bottom panel) a plaquette action.
}
\label{Wl-fig1}
\end{center}
\end{figure}

Let $S_g$ be a plaquette action for the $SU(N)$ lattice Yang-Mills theory:
\begin{align}
S_g  := & \sum_{n,\mu\ne\nu}  \frac{1}{g^2} 
{\rm tr}(U_{n,\mu}U_{n+\hat{\mu},\nu}U^{\dagger}_{n+\hat{\nu},\mu}U^{\dagger}_{n,\mu})
\nonumber\\
  = & \sum_{n,\mu<\nu}  \frac{1}{g^2}  {\rm tr}(U_{n,\mu\nu}+U^{\dagger}_{n,\mu\nu}) ,
\label{chap3-5}
\end{align}
where the link field $U_{n,\mu}$ satisfies $U_{n+\hat{\mu},-\mu} = U^{\dagger}_{n,\mu}$.
This action reproduces the ordinary Yang-Mills action $-\int d^Dx \sum_{\mu<\nu}{\rm tr} (F_{\mu\nu}^2)$
up to constant in the naive continuum limit (lattice spacing $\epsilon \to 0$). 
The diagrammatic expressions of a plaquette variable $U_{n,\mu\nu}$ and the plaquette action are given   in Fig.\ref{Wl-fig1}.


Note that the standard Wilson action $S_W$ is defined by 
\begin{align}
S_W = \sum_{n,\mu<\nu}  
\beta \left\{ \frac{1}{2{\rm tr}(\bf{1})}  {\rm tr}[U_{n,\mu\nu}+U^{\dagger}_{n,\mu\nu}] -1 \right\}, 
\label{wilson-action1}
\end{align}
see e.g., \cite{Creutz:text}. 
The difference of the constant term in the action is physically insignificant and we drop it in the strong coupling 
analysis.
By comparing $S_g$ and $S_W$, we can find 
\begin{align}
\beta=2N/g^2.
\label{wilson-action2}
\end{align}

We define a partition function $Z$ by
\begin{align}
Z   := \int\prod_{n,\mu}dU_{n,\mu} e^{S_g} ,
\label{chap3-4}
\end{align}
where $dU_{n,\mu}$ is the invariant integration measure of $SU(N)$. 
Then the expectation value $\langle W( C ) \rangle$ of  an operator $W( C )$  is defined by 
\begin{align}
 \langle W( C ) \rangle   := \frac{\int\prod_{n,\mu}dU_{n,\mu}e^{S_g} W( C )   }{Z} .
\label{chap3-3}
\end{align}

In order to evaluate the expectation value in eq.(\ref{chap3-3}), we perform the strong coupling expansion:  
For the large bare coupling constant $g$, we can expand the weight $e^{S_g}$ into the power-series of
$1/g^2$,
\begin{align}
e^{S_g} = \prod_{n,\mu<\nu} 
\left\{
\sum_{k=0}^{\infty} \frac{1}{k !} \left(\frac{1}{g^2}\right)^k [{\rm tr} (U_{n,\mu\nu})+{\rm tr} (U^{\dagger}_{n,\mu\nu})]^k
\right\},
\label{chap3-5-2}
\end{align}
and perform the group integration over each link variable $U_{n,\mu}$ according to the measure $dU_{n,\mu}$.
In Appendix A, we summarize the formulas needed for the strong coupling expansion and for the $SU(N)$ group integration.


\begin{figure}[tbp]
\begin{center}
\includegraphics[height=3.5cm]{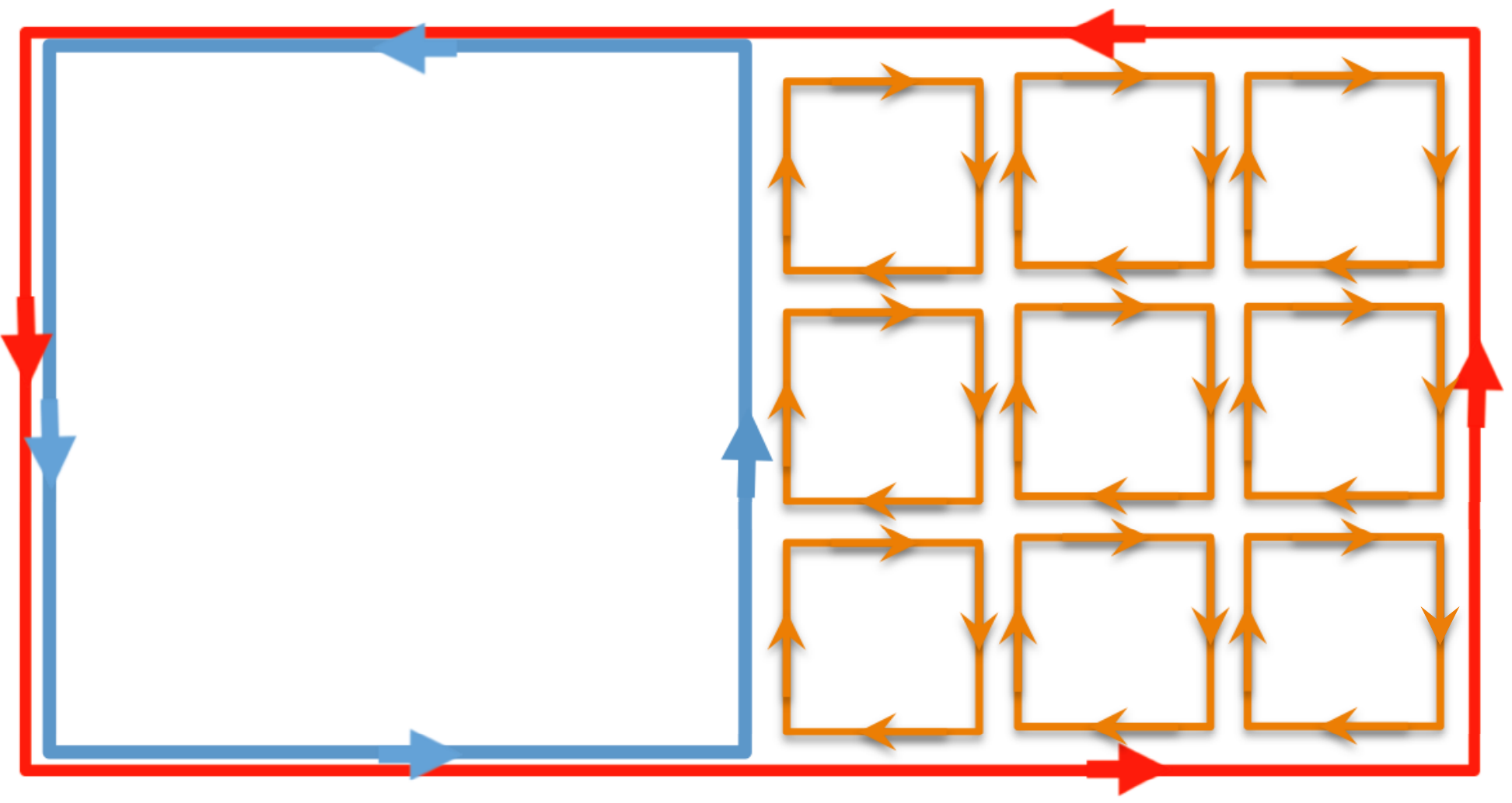}
\includegraphics[height=4.5cm]{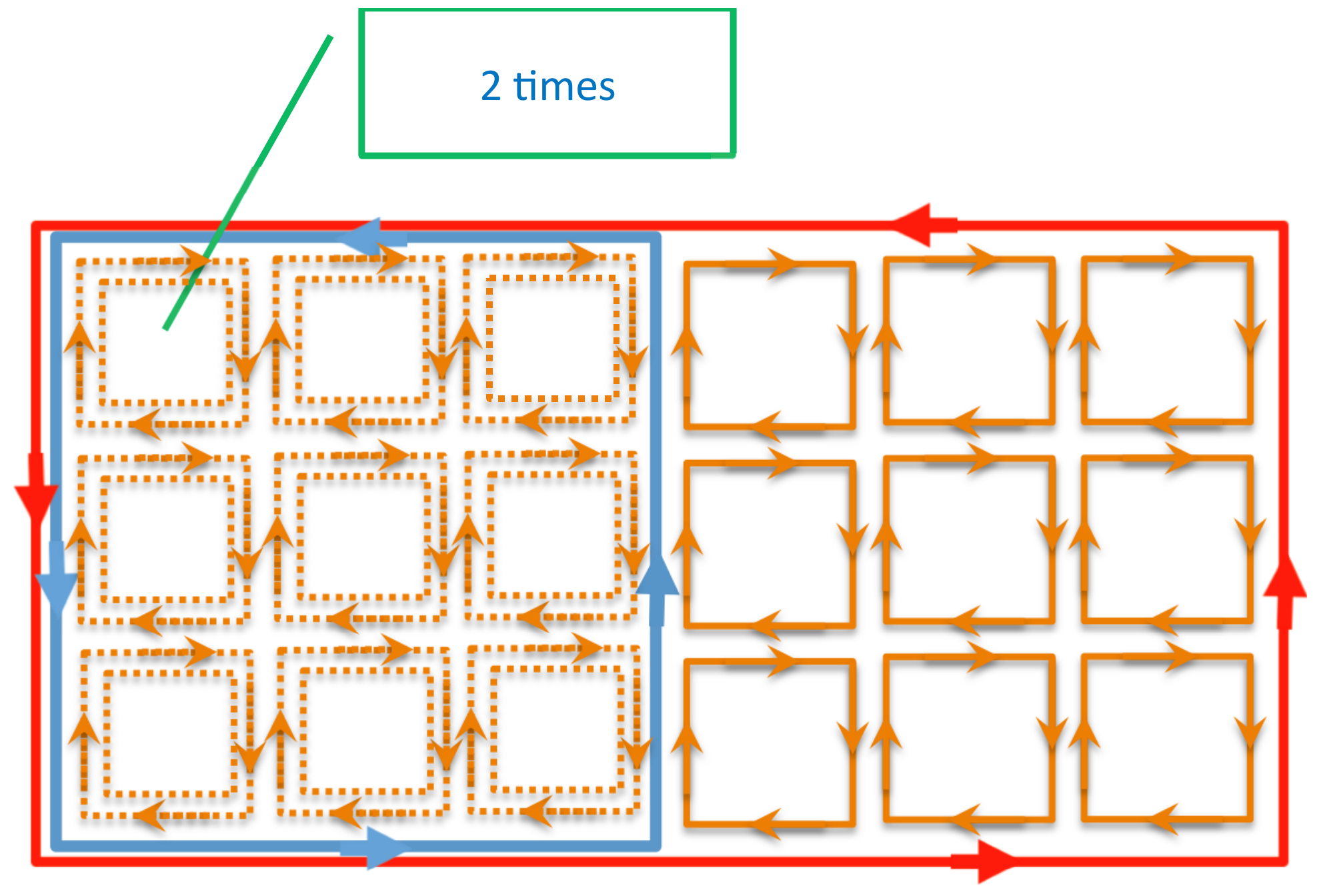}
\caption{ 
A set of plaquettes tiling the areas $S_1$ and $S_2$ which gives the non-trivial contribution to a coplanar double-winding Wilson loop average $\langle W(C_1\times C_2 ) \rangle$ for $SU(2)$.
(top panel) the leading contribution, 
(bottom panel) a higher order contribution. 
}
\label{Dw-fig6}
\end{center}
\end{figure}


\subsubsection{ $SU(2)$}

First, we study the case of $SU(2)$ gauge group.
For a coplanar double-winding Wilson loop, 
there is a single link variable $U_\ell$ for  a link $\ell \in C_1-C_2$ and there is a double link variable  $U_\ell U_\ell$ for a link $\ell \in  C_2$, 
as shown in the top diagram of Fig.\ref{Dw-fig6}.

We list some of explicit $SU(2)$  group integration formula as
\begin{subequations}
\begin{align}
& \int dU \   1 =1 ,
\label{2-sub2-5-1} 
\\
& \int dU \ U_{ab} =0 , \
  \int dU \ U^{\dagger}_{ab} =0 ,
\label{2-sub2-5-2b} \\
& \int dU \ U_{ab}U^{\dagger}_{kl} =\frac{1}{2}\delta_{al}\delta_{bk} ,
\label{2-sub2-5-3} 
\\
& \int dU \ U_{a_1b_1}U_{a_2b_2} =\frac{1}{2!}\epsilon_{a_1a_2 }\epsilon_{b_1b_2 } 
%
= \int dU \ U^{\dagger}_{a_1b_1}U^{\dagger}_{a_2b_2} ,
\label{2-sub2-5-4}
\\
& \int dU \ U_{a_1b_1}U_{a_2b_2}\cdots U_{a_Mb_M} =0, \ M\ne 0 \ ({\rm mod} \ 2),
\label{2-sub2-5-5}
\\ 
&  \int dU \ U_{ab}U_{cd}U^{\dagger}_{ij}U^{\dagger}_{kl}  
\nonumber \\ 
=& 
\frac{1}{(2^2-1)}[
\delta_{aj}\delta_{bi}\delta_{cl}\delta_{dk}+\delta_{al}\delta_{bk}\delta_{cj}\delta_{di}
\nonumber \\&
-\frac{1}{2}(
\delta_{aj}\delta_{bk}\delta_{cl}\delta_{di}+\delta_{al}\delta_{bi}\delta_{cj}\delta_{dk}
)]   
 +\left(  \frac{1}{2!}\right)^{2}\epsilon_{ac}\epsilon_{bd}\epsilon_{ik}\epsilon_{jl}.
\label{2-sub2-5-6b}
\end{align}
\end{subequations}

For a single link variable $U_\ell$ (resp. $U^{\dagger}_\ell$) for $\ell \in C_1-C_2$, we need at least one additional link variable with an opposite direction 
$U^{\dagger}_\ell$ (resp. $U_\ell$) to obtain non-vanishing result after integration in eq.(\ref{chap3-3})  
according to the integration formulas (\ref{2-sub2-5-3}) for the $SU(2)$ group integrations. 
Such link variables are supplied from the expansion eq.(\ref{chap3-5-2}) of $e^{S_g}$. 
Since the number of plaquettes which are brought down from $e^{S_g}$ must be equal to the power of $1/g^2$ in the expansion eq.(\ref{chap3-5-2}),
the leading contribution to $\langle W(C_1\times C_2 ) \rangle$ comes from a set of plaquettes  tiling the minimal area $S_1-S_2$ with the least number of plaquettes.
See the top diagram of Fig.\ref{Dw-fig6}.
For double link variables $U_\ell U_\ell$ for $\ell \in  C_2$, on the other hand, we do not need additional link variables coming from the expansion of $e^{S_g}$ to obtain the non-vanishing result due to the integration (\ref{2-sub2-5-4}), giving the $g$-independent contribution.

For the $SU(2)$ gauge group, therefore, 
the leading contribution to $\langle W(C_1\times C_2 ) \rangle$ in the strong coupling expansion  comes from the term in which a set of plaquettes tiles the surface with the area $S_1-S_2$, as shown in the top diagram of Fig.\ref{Dw-fig6}.
Therefore, group integrations give the result 
\begin{align}
\langle W(C_1\times C_2 ) \rangle_{\rm leading} = -2\left(   \frac{1}{2g^2} \right)^{S_1-S_2} 
= -2e^{-\sigma (S_1-S_2)},
\label{sce-su2-1}
\end{align}
where $\sigma=\log(2g^2)$.
This result was first obtained by Greensite and H\"ollwieser in \cite{GH15}.
We reconfirmed the difference-of-areas law of coplanar double-winding Wilson loops
for $SU(2)$.
The  bottom diagram of Fig.\ref{Dw-fig6} shows one of  higher-order contributions in the strong coupling expansion for $SU(2)$.
This diagram gives non-vanishing contribution due to the integration formula (\ref{2-sub2-5-6b}). 

\subsubsection{ $SU(N)$, ($N\geq 3$)}

Next, we study the case of  $SU(N)$ ($N\geq 3$)
gauge groups.  
We list some of explicit $SU(N)$ ($N\geq 3$) group integration formula as
\begin{subequations}
\begin{align}
& \int dU \ 1 =1,
\label{sub2-5-1} 
\\
& \int dU \ U_{ab} =0,
\label{sub2-5-2}
\\
& \int dU \ U_{ab}U^{\dagger}_{kl} =\frac{1}{N}\delta_{al}\delta_{bk},
\label{sub2-5-3} 
\\
& \int dU \ U_{a_1b_1}U_{a_2b_2}\cdots U_{a_Mb_M} =0, \  M\ne 0 \ ({\rm mod}\ N),
\label{sub2-5-5}
\\
& \int dU \ U_{a_1b_1}U_{a_2b_2}\cdots U_{a_Nb_N} =\frac{1}{N!}\epsilon_{a_1a_2\cdots a_N}\epsilon_{b_1b_2\cdots b_N},
\label{sub2-5-4} 
\\
& \int dU \ U_{ab}U_{cd}U^{\dagger}_{ij}U^{\dagger}_{kl} 
\nonumber\\&
=
\frac{1}{(N^2-1)}[
\delta_{aj}\delta_{bi}\delta_{cl}\delta_{dk}+\delta_{al}\delta_{bk}\delta_{cj}\delta_{di}
\nonumber\\&
-\frac{1}{N}(
\delta_{aj}\delta_{bk}\delta_{cl}\delta_{di}+\delta_{al}\delta_{bi}\delta_{cj}\delta_{dk}
)] .
\label{sub2-5-6}
\end{align}
\end{subequations}


\begin{figure}[tbp]
\begin{center}
\includegraphics[height=5.0cm]{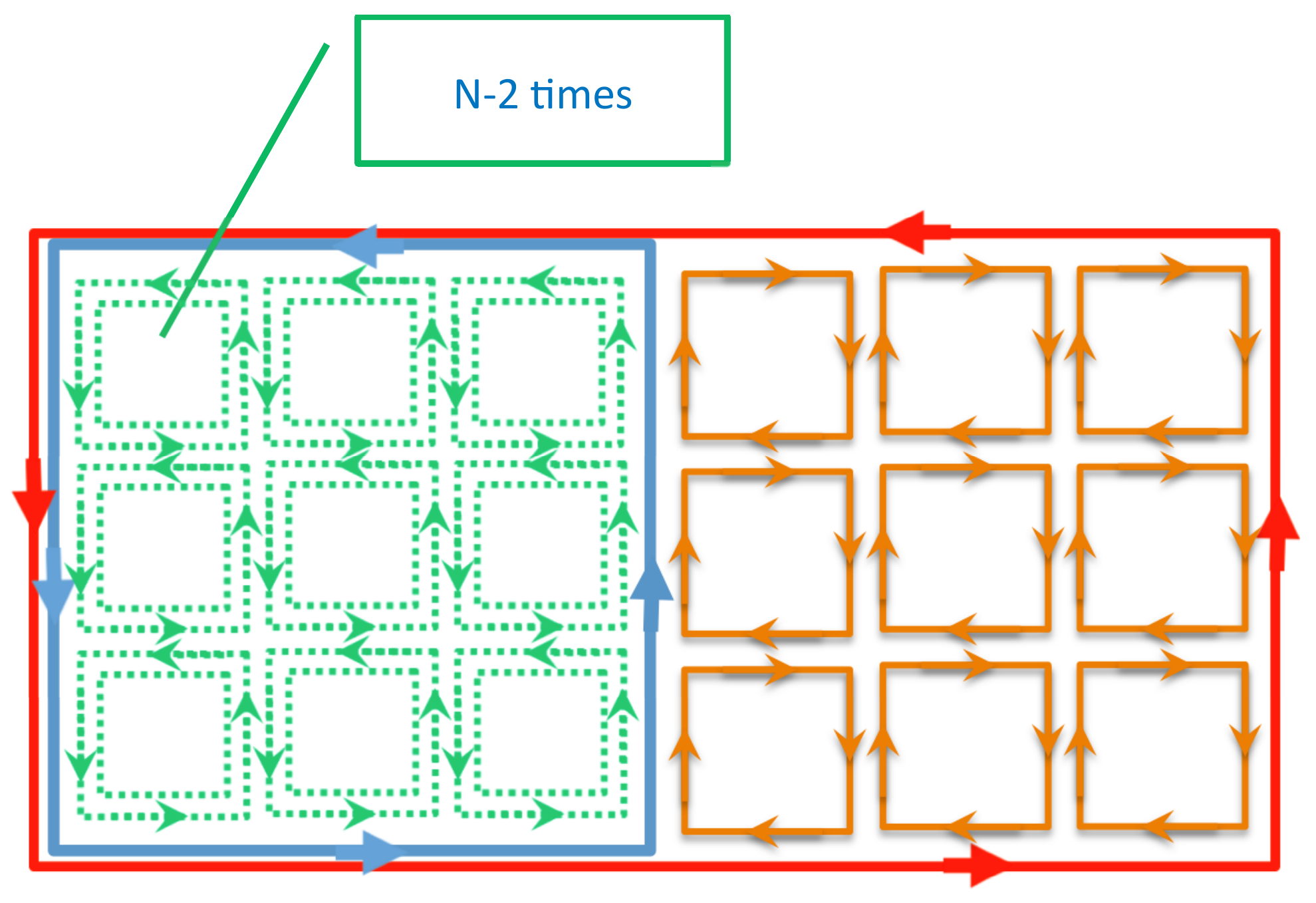}
\includegraphics[height=5.0cm]{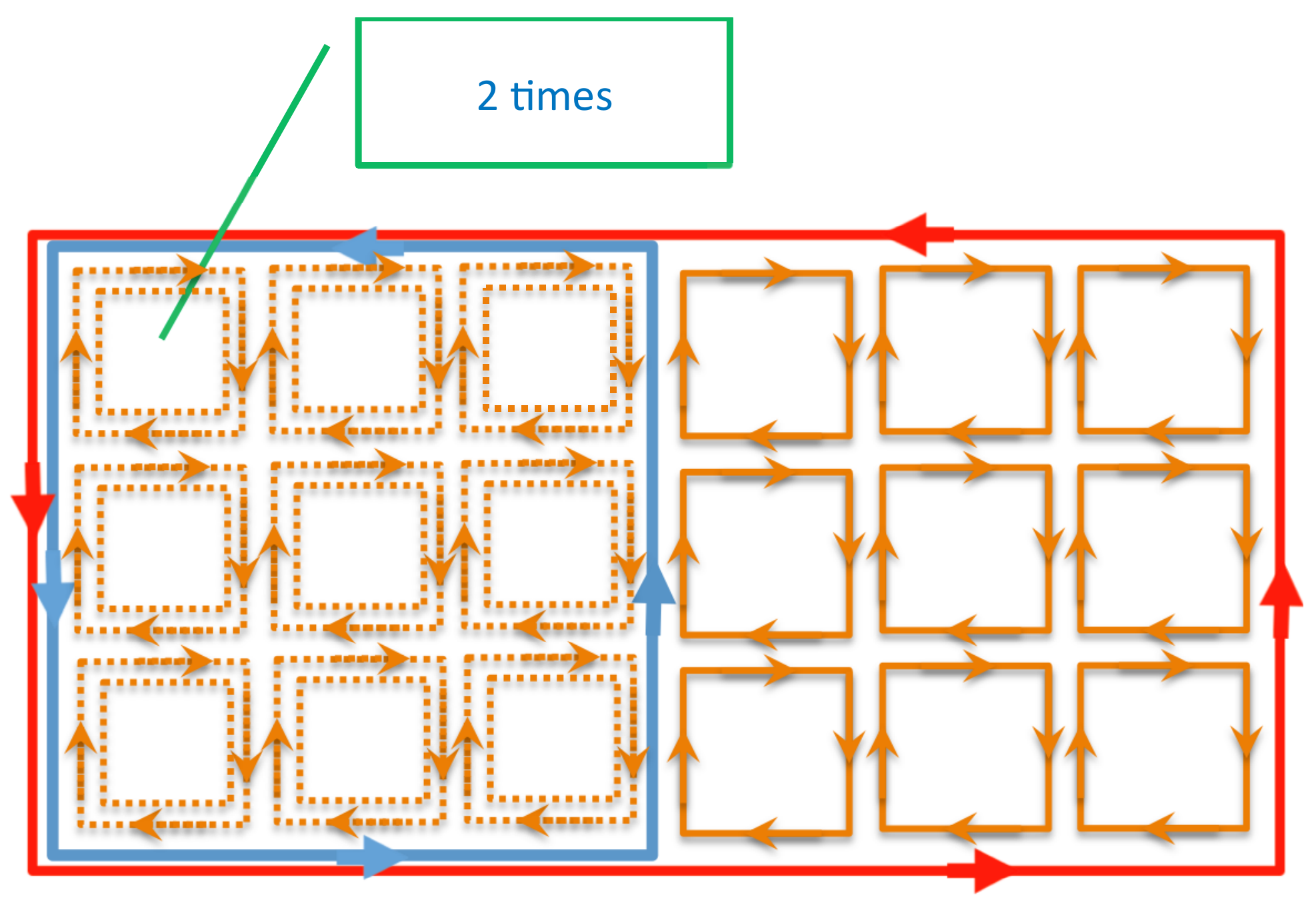}
\caption{ 
A set of plaquettes tiling the areas $S_1$ and $S_2$ which gives the leading contribution to a coplanar double-winding Wilson loop average $\langle W(C_1\times C_2 ) \rangle$ for $SU(N)$ ($N\geq 3$).
Here $S_1(= L \times L_2)$ and $S_2(= L_1\times L_2)$ are respectively the minimal areas bounded by rectangular loops $C_1$ and $C_2$ with $S_1\geq S_2$.
For $N=3$, the diagram of the top panel gives the leading contribution and that of the bottom panel gives the next-to-leading contribution. 
For $N=4$, the two diagrams give identical contributions. 
For $N>4$, the diagram of the bottom panel gives the leading contribution and that of the top panel gives the next-to-leading contribution. 
}
\label{Dw-fig7}
\end{center}
\end{figure}


Notice that the $SU(N)$ case is different from the $SU(2)$ case. 
For a double link variable  $U_\ell U_\ell$ for a link $\ell \in  C_2$, we need additional $N-2$ link variables $(U_\ell)^{N-2}$ with the same direction to be brought down from the expansion of $e^{S_g}$  in eq.(\ref{chap3-3}) to obtain the non-vanishing result after the integration according to the integration formulas (\ref{sub2-5-4}) for the $SU(N)$ group integrations. 
See the top diagram of Fig.\ref{Dw-fig7}.
For a single link variable $U_\ell$ (resp. $U^{\dagger}_\ell$) for a link $\ell \in C_1-C_2$, on the other hand, we need at least one additional link variable with the opposite direction $U^{\dagger}_\ell$ (resp. $U_\ell$) to obtain non-vanishing result after integration in eq.(\ref{chap3-3}) according to the integration formulas (\ref{sub2-5-3}) for the $SU(2)$ group integrations.  
Therefore, the contribution from the top diagram of Fig.\ref{Dw-fig7} is given by 
\begin{align}
p_N \left(  \frac{1}{g^2N} \right)^{(N-2)S_2+(S_1-S_2)} ,
\end{align}
where the coefficient $p_N$ is calculated by collecting the numerical factors coming from link integrations and  the power-series expansions of $e^{S_g}$.

We have another contribution from the bottom diagram of Fig.\ref{Dw-fig7}.
For a double link variable  $U_\ell U_\ell$ with the same direction for a link $\ell \in  C_2$, we have additional $2$ link variables $(U_\ell^\dagger)  (U_\ell^\dagger)$ with the opposite directions to be brought down from the expansion of $e^{S_g}$ in eq.(\ref{chap3-3}) to obtain the non-vanishing result after the integration according to the integration formulas (\ref{sub2-5-6}) for the $SU(N)$ group integrations.
For a single link variable $U_\ell$ (resp. $U^{\dagger}_\ell$) for a link $\ell \in C_1-C_2$, on the other hand, we need at least one additional link variable with an opposite direction 
$U^{\dagger}_\ell$ (resp. $U_\ell$) to obtain non-vanishing result after integration in eq.(\ref{chap3-3})  
according to the integration formulas (\ref{sub2-5-3}) for the $SU(N)$ group integrations.  
Therefore, the contribution from the bottom diagram of Fig.\ref{Dw-fig7} is given by 
\begin{align}
q_N \left(  \frac{1}{g^2N} \right)^{2S_2+(S_1-S_2)} =
q_N \left(  \frac{1}{g^2N} \right)^{S_1+S_2} ,
\end{align}
where the coefficient  $q_N$ is calculated in the similar way to $p_N$.  

For the $SU(N)$ ($N\geq 3$), the leading contribution in the strong coupling expansion may 
come  from one of the two diagrams shown in Fig.\ref{Dw-fig7}. 
Since the number of plaquettes brought down from $e^{S_g}$ is equal  to the power of  $1/g^2$,
these two contributions can be written as
\begin{align}
\langle W(C_1\times C_2 ) \rangle =&
p_N \left(  \frac{1}{g^2N} \right)^{(N-2)S_2+S_1-S_2}
\nonumber\\&
+q_N \left(  \frac{1}{g^2N} \right)^{S_1+S_2} + \cdots,
\label{sce-suN-1}
\end{align}
where coefficients $p_N$, $q_N$ are determined by expansion coefficients of the power series expansion of $e^{S_g}$ and $SU(N)$ group integrations for link variables.
Which contribution becomes dominant is naively determined by comparing the power index of $\frac{1}{g^2N}$, which depends on the number of color $N$.

For $N\geq4$, we find that the second term in eq.(\ref{sce-suN-1}) gives the dominant contribution in the strong coupling expansion    
for $\langle W(C_1\times C_2 ) \rangle$, since the inequality holds, 
$S_1+S_2 \leq (N-2)S_2+S_1-S_2$ for $N\geq4$. 
Thus we conclude that the sum-of-areas law of a coplanar double-winding Wilson loop is allowed for $N\geq 4$. 
This result is consistent with the result obtained by Matsudo and Kondo  in \cite{matsudo-kondo}.

From the top panel of Fig.\ref{Dw-fig7}, we can easily find that the coefficient $p_N$ should be calculated for 
each number of color $N$, because type of diagrams are different with the number of color $N$. 
On the other hands,  we can obtain general formula for the coefficient $q_N$, since the diagram of the bottom panel of Fig.\ref{Dw-fig7}  is common to all numbers of color $N$. The result is
\begin{align}
 q_N =  -\frac{N^{2S_2}}{2} 
\left\{ 
\left[ \frac{1}{N(N-1)}\right]^{S_2-1} - \left[   \frac{1}{N(N+1)} \right]^{S_2-1}
\right\},
\label{sce-q_N}
\end{align}
for $S_2 \geq 1$ in lattice units.
See Appendix B for the detail.

In the following, we show the results for $SU(2)$, $SU(3)$ and $SU(4)$ in more detail.

\vspace{0.5cm}
\underline{ $SU(2)$}
For the number of color $N=2$,  eq. (\ref{sce-suN-1}) reduces to 
\begin{align}
\langle W(C_1\times C_2 ) \rangle =
2p_2 \left(  \frac{1}{2g^2} \right)^{S_1-S_2}
+2q_2 \left(  \frac{1}{2g^2} \right)^{S_1+S_2} + \cdots ,
\label{sce-su2r-1}
\end{align}
where
\begin{align}
 p_2 =& -2,  \\
 q_2 =&  -\frac{4^{S_2}}{2} 
\left\{ 
\left[ \frac{1}{2}\right]^{S_2-1} - \left[   \frac{1}{6} \right]^{S_2-1}
\right\},  \ (S_2 \geq 1).
\label{sce-su2r-2}
\end{align}

The factor $2$ in front of $p_2$ and $q_2$ arises from the non-oriented nature of the plaquettes 
for $SU(2)$, which is to be compared with (\ref{sce-su2-1}).

\vspace{0.5cm}
\underline{ $SU(3)$}
For the number of color $N=3$,  eq. (\ref{sce-suN-1}) reduces to 
\begin{align}
\langle W(C_1\times C_2 ) \rangle =
p_3 \left(  \frac{1}{3g^2} \right)^{S_1}
+q_3 \left(  \frac{1}{3g^2} \right)^{S_1+S_2} + \cdots ,
\label{sce-su3-1}
\end{align}
where
\begin{align}
 p_3 =& -3,  \\
 q_3 =&  -\frac{9^{S_2}}{2} 
\left\{ 
\left[ \frac{1}{6}\right]^{S_2-1} - \left[   \frac{1}{12} \right]^{S_2-1}
\right\},  \ (S_2 \geq 1).
\label{sce-su3-2}
\end{align}

The coefficient $q_3$ is obtained from eq.(\ref{sce-q_N}). See Appendix C for the calculation of $p_3$.

From this result, we find that the first term in eq.(\ref{sce-su3-1}) gives the dominant contribution to $\langle W(C_1\times C_2 ) \rangle$ for 
sufficiently large areas $S_1$ and $S_2$, which is neither difference-of-areas law nor sum-of-areas law for
the area-law falloff of the coplanar double-winding Wilson loop average. We call this area-law falloff ``max-of-areas law'' (or $\max(S_1,S_2)$ law).
This result is also consistent with the result obtained by Matsudo and Kondo  in \cite{matsudo-kondo}.

\vspace{0.5cm}
\underline{ $SU(4)$}
For the number of color $N=4$, 
eq. (\ref{sce-suN-1}) reduces to 
\begin{align}
\langle W(C_1\times C_2 ) \rangle =
p_4 \left(  \frac{1}{4g^2} \right)^{S_1+S_2}
+q_4 \left(  \frac{1}{4g^2} \right)^{S_1+S_2} + \cdots,
\label{sce-su4-1}
\end{align}
where
\begin{align}
 p_4 =&  -8  \left[   \frac{1}{12} \right]^{S_2-1} ,\\
 q_4 =&   -\frac{16^{S_2}}{2} 
\left\{ 
\left[ \frac{1}{12}\right]^{S_2-1} - \left[   \frac{1}{20} \right]^{S_2-1}
\right\},  \ (S_2 \geq 1).
\label{sce-su4-2}
\end{align}
In this case, both terms in eq.(\ref{sce-su4-1}) behave as sum-of-areas law.


\subsubsection{ $L_1$ dependence of the $\langle W(C_1\times C_2 ) \rangle$}

\begin{figure}[tbp]
\begin{center}
\includegraphics[height=4.5cm]{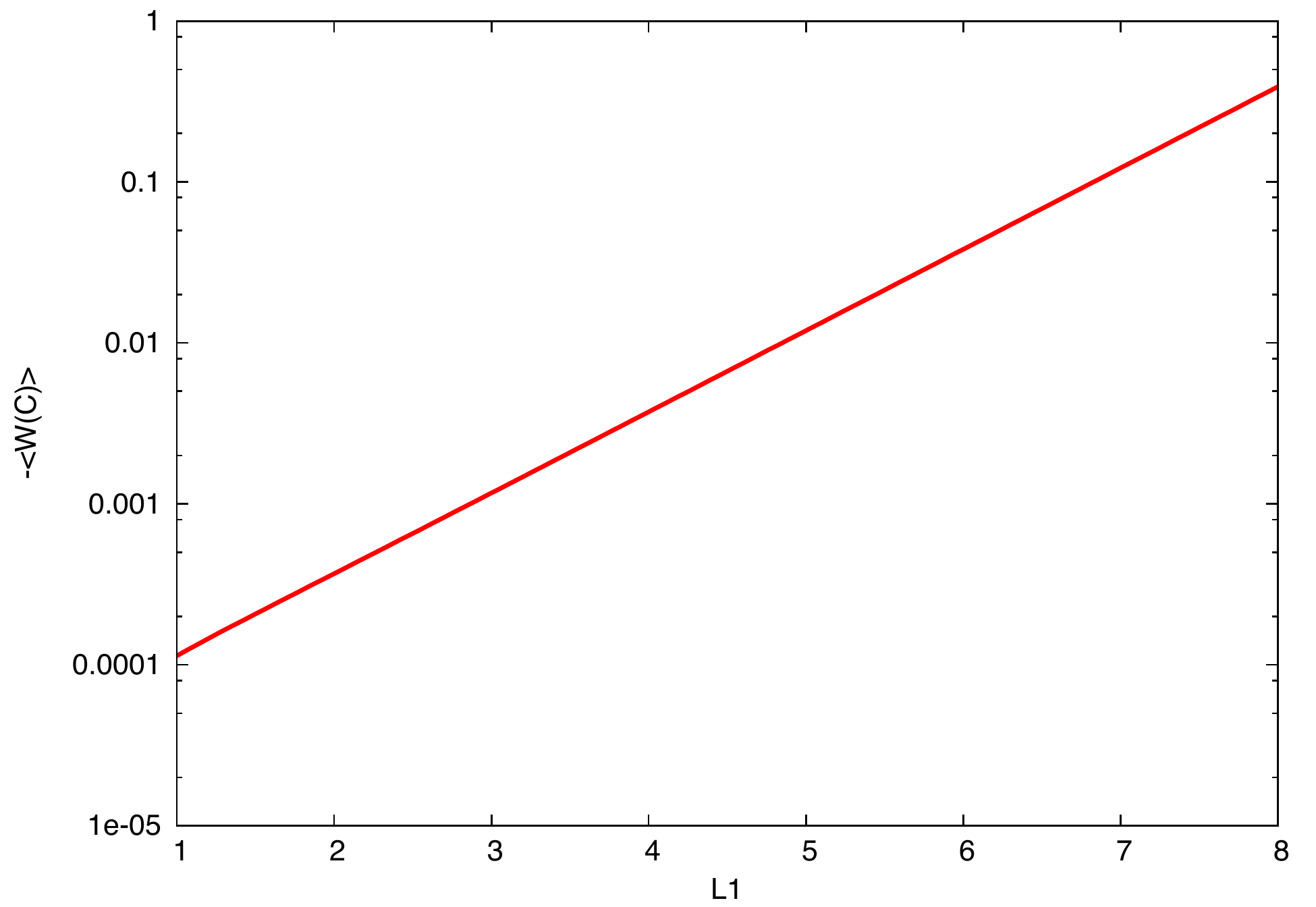}
\caption{
$L_1$-dependence of a coplanar double-winding Wilson loop average  $- \langle W(C_1\times C_2 ) \rangle$ from the strong coupling expansion in $SU(2)$ lattice gauge theory.
We plot eq.(\ref{sce-su2r-1}) times $-1$ versus $L_1=1 \sim 10$ for $L=10$, $L_2=1$ and $1/g^2N=2.5/8$.
}
\label{numerical-fig1}
\end{center}
\end{figure}


\begin{figure}[tbp]
\begin{center}
\includegraphics[height=5.0cm]{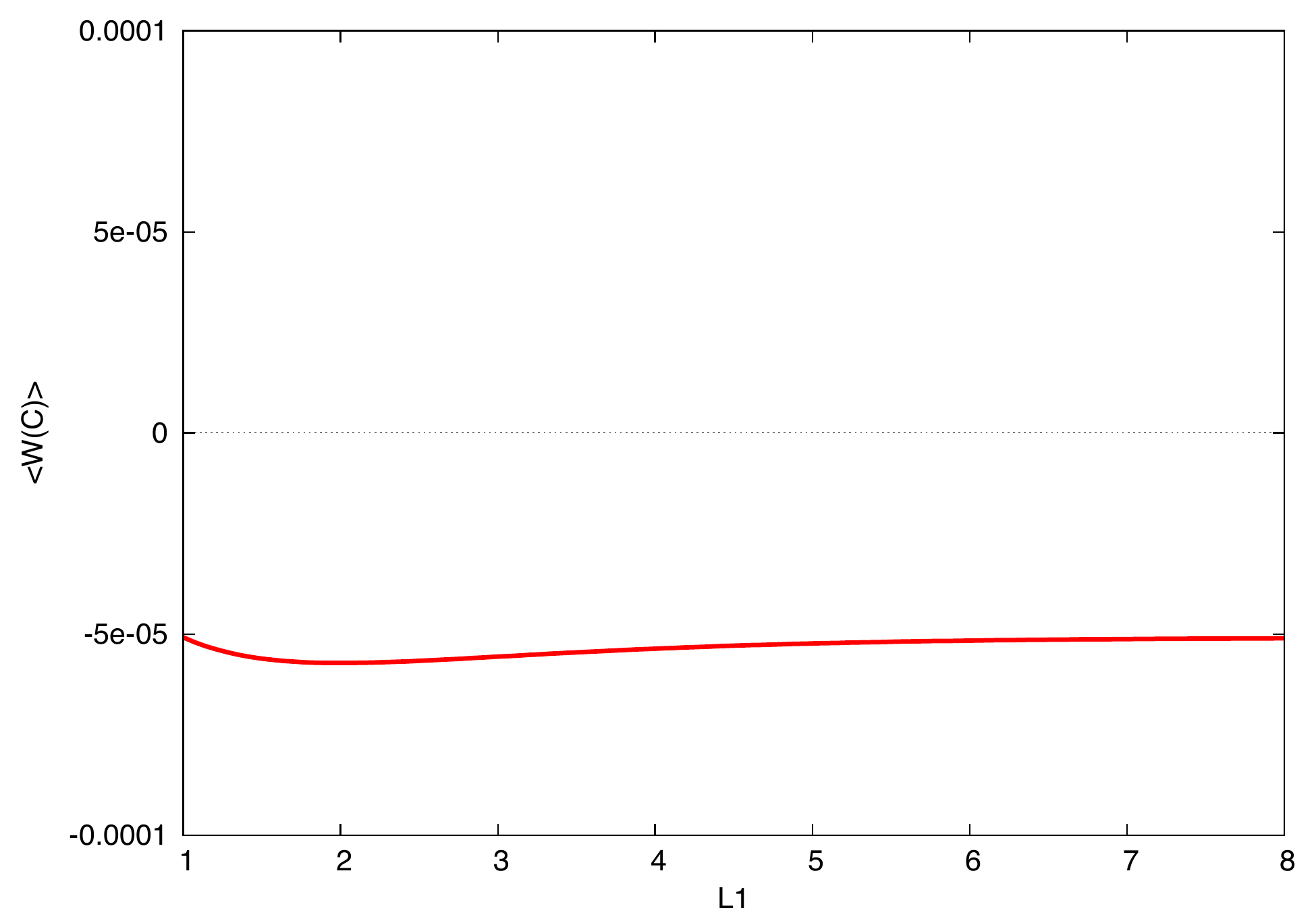}
\includegraphics[height=5.0cm]{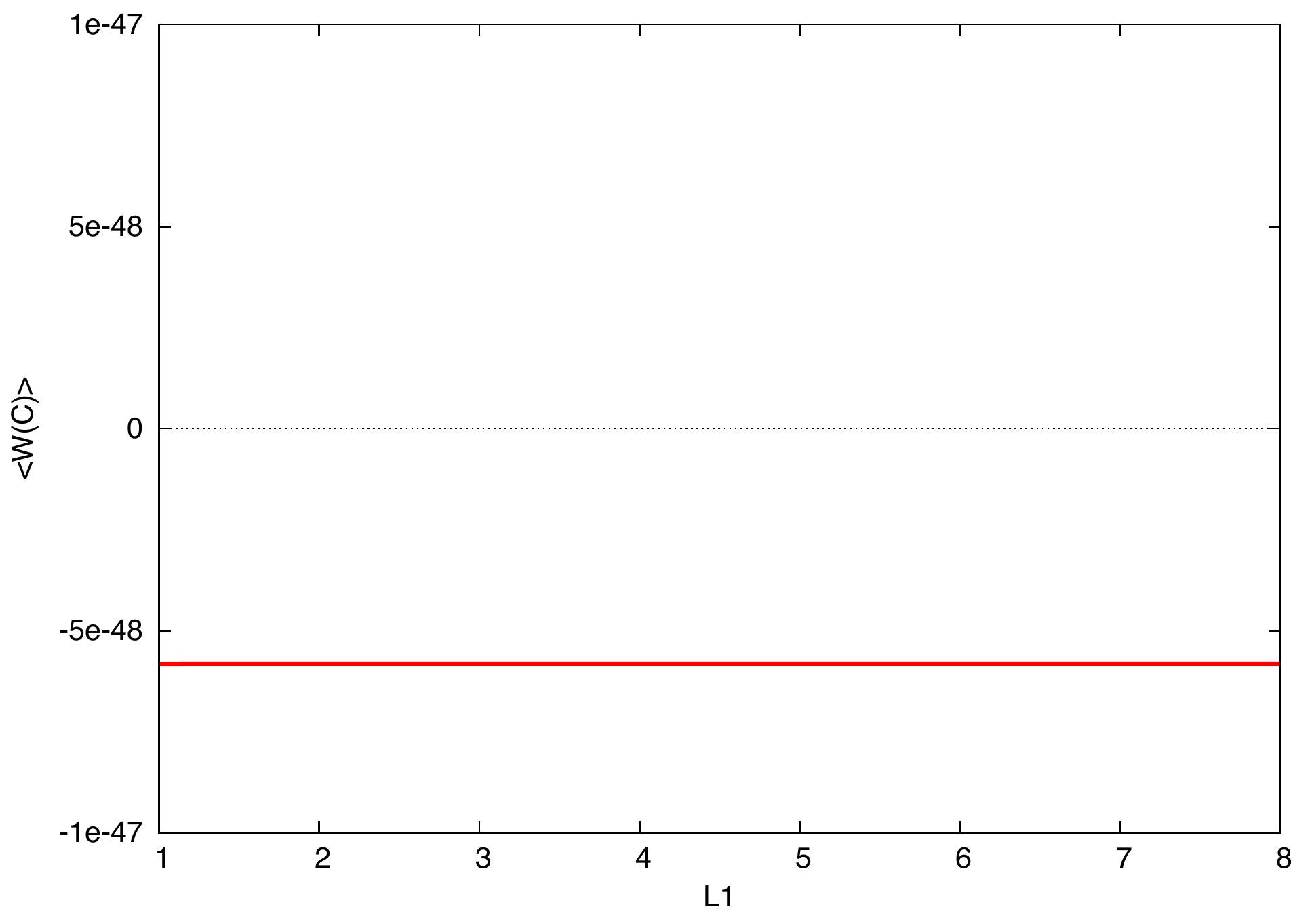}
\caption{ 
$L_1$-dependence of a coplanar double-winding Wilson loop average  $\langle W(C_1\times C_2 ) \rangle$ from the strong coupling expansion in $SU(3)$ lattice gauge theory.
We plot eq.(\ref{sce-su3-1}) versus $L_1=1 \sim 8$ for  $1/g^2N=6.0/18$.
(top panel) $L=10$, $L_2=1$.
(bottom panel)  $L=10$, $L_2=10$.
}
\label{numerical-fig2}
\end{center}
\end{figure}


From the above discussions, we can understand the $L_1$ dependence of the coplanar double-winding Wilson loop average 
$\langle W(C_1\times C_2 ) \rangle$ in $SU(N)$ lattice Yang-Mills gauge theory for fixed $L$, $L_2$, and gauge coupling $g$.

For $SU(2)$ gauge group, we plot eq.(\ref{sce-su2r-1}) in Fig.\ref{numerical-fig1}, which shows the difference-of-areas law behavior of a coplanar double-winding Wilson loop for $N=2$.

On the other hand, we plot eq.(\ref{sce-su3-1}) in Fig.\ref{numerical-fig2}. For $SU(3)$ gauge group, as the coplanar double-winding Wilson loop average follows the  max-of-areas law, it is expected that there are no $L_1$-dependence  of  $\langle W(C_1\times C_2 ) \rangle$ for efficiently large areas $S_1$ and $S_2$. In fact, we can see that the plots flatten  at  $L_1\sim 4$ (resp. $L_1\sim 1$) in top (resp. bottom) panel in Fig.\ref{numerical-fig2}.

\subsection{Numerical simulation}

We examine the $L_1$-dependence of $\langle W(C_1\times C_2 ) \rangle$ that we discussed above.

\underline{$SU(2)$}:
We generate the configurations of $SU(2)$ link variables $\{U_{n,\mu}\}$, using the (pseudo-)heat-bath method for the standard Wilson  action. The numerical simulations are performed on the $24^4$ lattice at $\beta(=2N/g^2)=2.5$. We thermalize $3000$ sweeps, and in particular, we have used $100$ configurations for calculating the expectation value of coplanar double-winding Wilson loops $\langle W(C_1\times C_2 ) \rangle$.

\begin{figure}[tbp]
\begin{center}
\includegraphics[height=5.0cm]{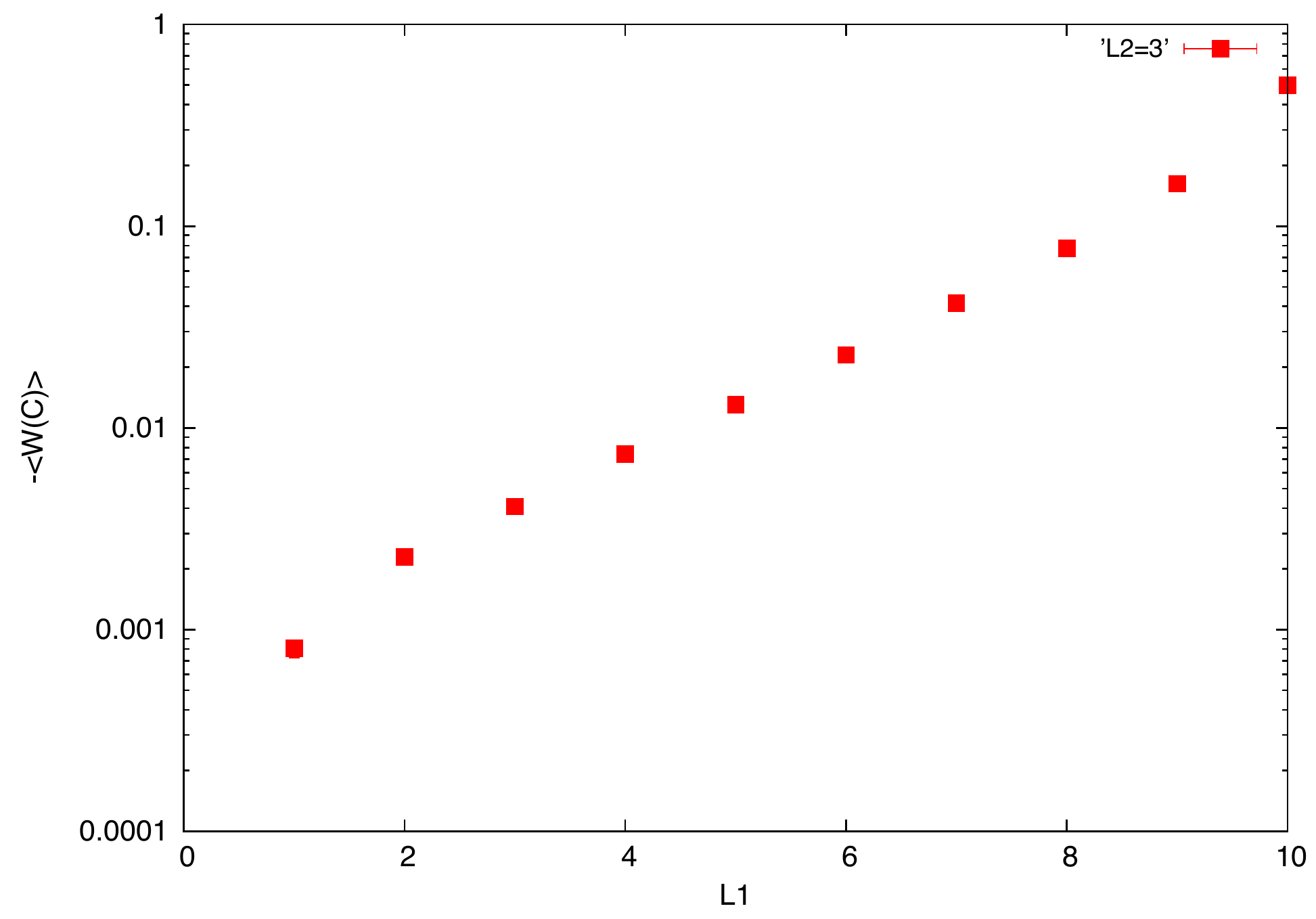}
\caption{ 
$L_1$-dependence of a coplanar double-winding Wilson loop average $-\langle W(C_1\times C_2 ) \rangle$ in the $SU(2)$ lattice gauge theory obtained from numerical simulations on a lattice of size $24^4$ at $\beta=2.5$ for fixed $L=10$, and $L_2=3$.
}
\label{numerical-fig3}
\end{center}
\end{figure}

Fig.\ref{numerical-fig3} shows the obtained plot for the $- \langle W(C_1\times C_2 ) \rangle$ for various value of $L_1$, when we choose parameters $L=10$, $L_2=3$.
The results of numerical simulations are consistent with analytical results in Fig.\ref{numerical-fig1}.
Thus we reconfirm the difference-of-areas law for $SU(2)$.
Note that we can also confirm $\langle W(C_1\times C_2 ) \rangle\simeq -1/2$ for $S_1=S_2$ from Fig.\ref{numerical-fig1}.

\underline{$SU(3)$}:
We also generate the configurations of $SU(3)$ link variables $\{U_{n,\mu}\}$, using the (pseudo-)heat-bath 
method for the standard Wilson  action.  The numerical simulations are performed on the $24^4$ lattice
at $\beta=6.2$. We have used $200$ configurations for calculating the expectation value of coplanar 
double-winding Wilson loops $\langle W(C_1\times C_2 ) \rangle$, where we have used APE smearing method ($N=12$, 
$\alpha=0.1$) as a noise reduction technique. See \cite{shibata} for the detail.

\begin{figure}[tbp]
\begin{center}
\includegraphics[height=5.0cm]{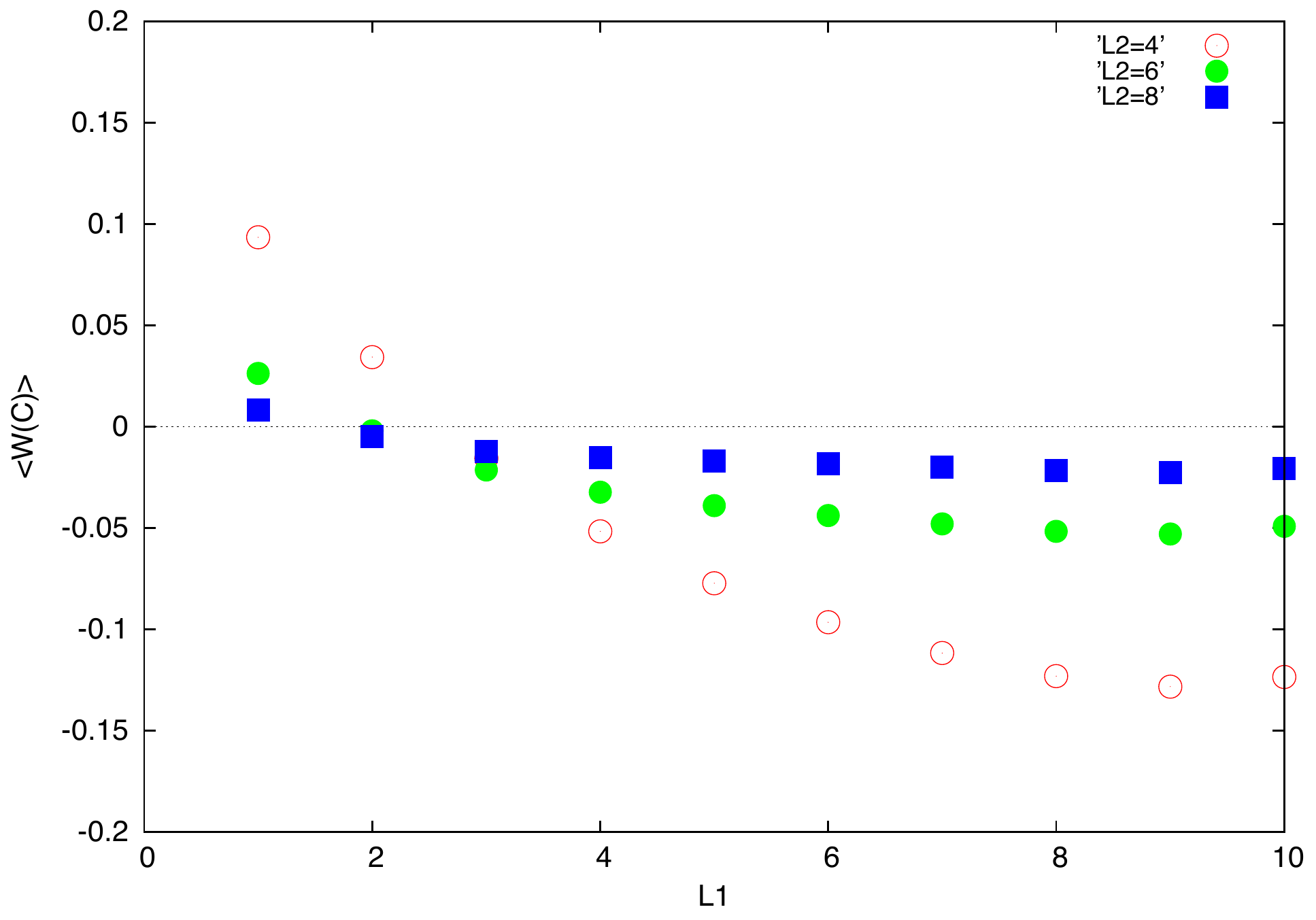}
\caption{ 
$L_1$-dependence of a coplanar double-winding Wilson loop average $\langle W(C_1\times C_2 ) \rangle$ in $SU(3)$ lattice gauge theory obtained from numerical simulations on a lattice of size $24^4$ at $\beta=6.2$ for fixed $L=10$, and $L_2=4,6,8$.
}
\label{numerical-fig4}
\end{center}
\end{figure}

Fig.\ref{numerical-fig4} shows the obtained plot for the $\langle W(C_1\times C_2 ) \rangle$ for various value of $L_1$,
when we choose parameters $L=10$, $L_2=4,6,8$.
The results of numerical simulations are consistent with analytical results in Fig.\ref{numerical-fig2}.
For example, we can see that the plots flatten at $L_1\sim 4$ for $L_2=8$, which means that there are no $L_1$-dependence  
of  $\langle W(C_1\times C_2 ) \rangle$.
Thus, we numerically confirm the max-of-areas law for $SU(3)$.

\section{A "shifted" double-winding Wilson loops}

\begin{figure}[tbp]
\begin{center}
\includegraphics[height=4.0cm]{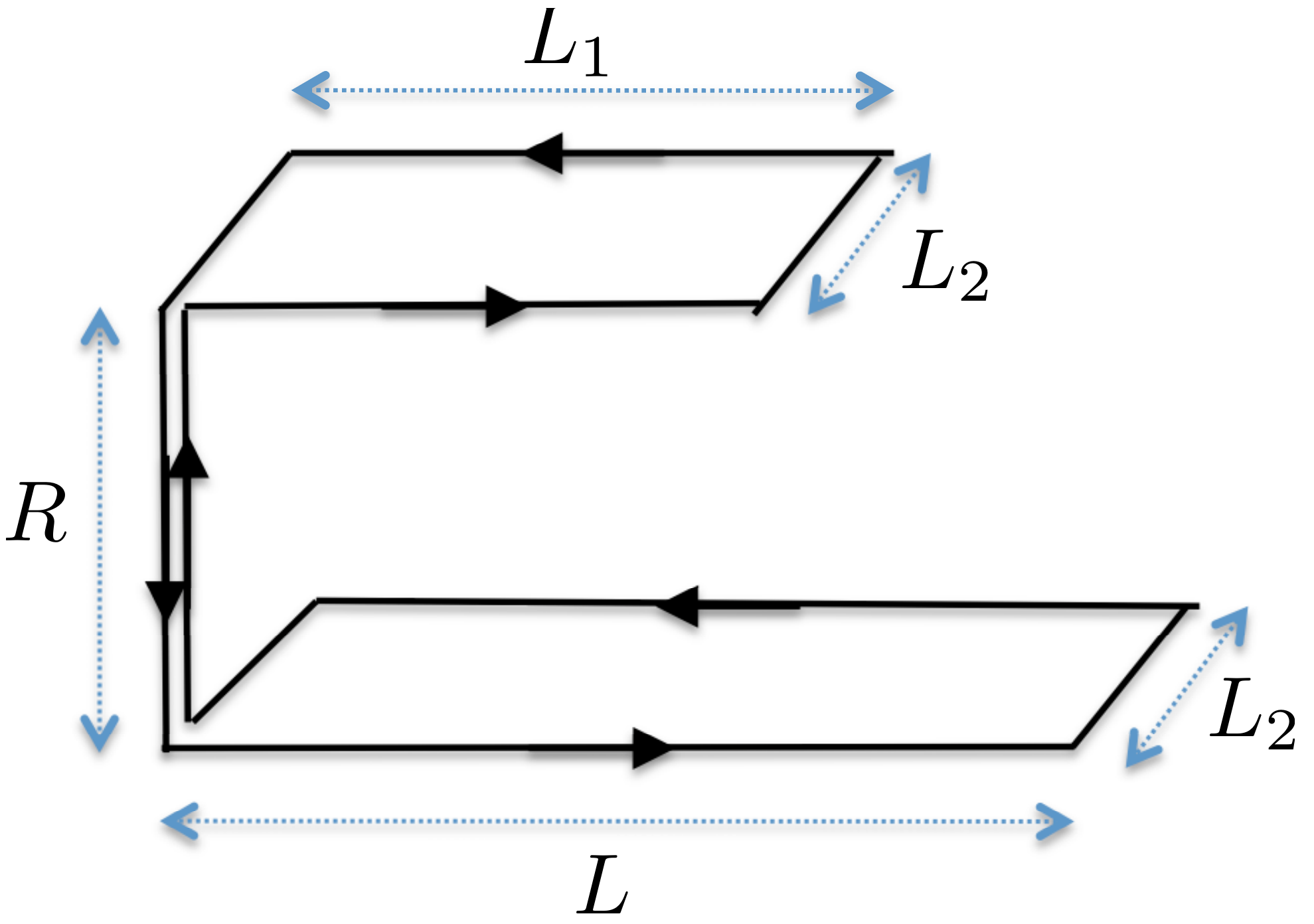}
\caption{  
The setting up of a shifted double-winding Wilson loop operator $W(C_1\times C_2 ) _{R \ne 0}$.
}
\label{Dw-fig8}
\end{center}
\end{figure}

Finally, we consider the shifted case $R \not= 0$ of a double-winding Wilson loop  in the $SU(N)$ lattice Yang-Mills gauge theory, as indicated in Fig.\ref{Dw-fig8}.
Contours $C_1$ and $C_2$ lie in planes parallel to the $x$-$t$ plane, but are displaced from one another in the $z$ direction by distance $R$.
Just like the previous section, for simplicity,  let $C_1$ ($C_2$) be a rectangular loop of length $L$, $L_2$ ($L_1$, $L_2$), and $S_1(\equiv L\times L_2)$,  $S_2(\equiv L_1\times L_2)$ be the minimal areas of contour $C_1$, $C_2$ respectively.

\subsection{strong coupling expansion}

First, we study the shifted double-winding Wilson loop  based on the strong coupling expansion.

\begin{figure}[tbp]
\begin{center}
\includegraphics[height=2.5cm]{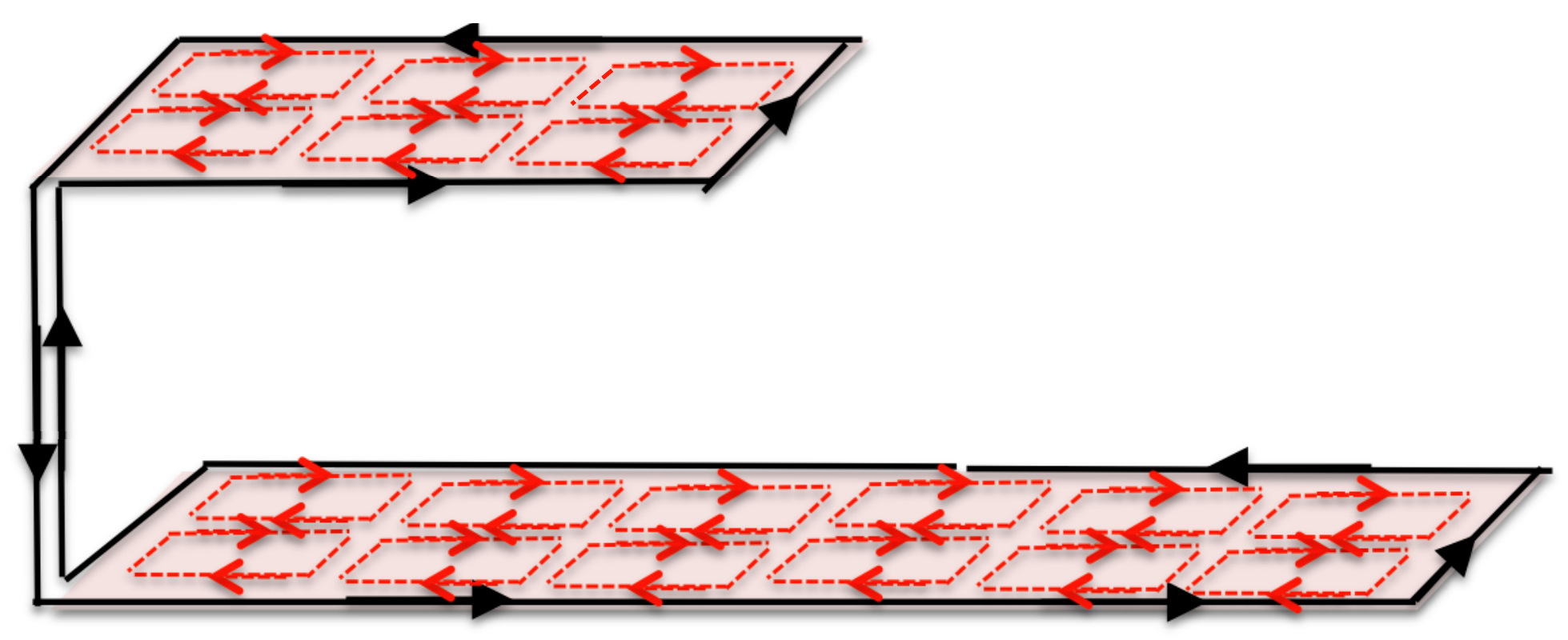}
\caption{  
One of diagrams which also contributes to a shifted double-winding Wilson loop average $\langle W(C_1\times C_2 ) \rangle_{R \ne 0}$ in the strong coupling expansion  of the $SU(N)$ lattice Yang-Mills theory.
}
\label{Dw-fig10}
\end{center}
\end{figure}
One of the diagrams which gives a leading contribution in the strong coupling expansion is given by a set of plaquettes tiling the two minimal surfaces $S_1$ and $S_2$, as shown in Fig.\ref{Dw-fig10}.
The results of a group integration for the links $U_\ell$'s on both surfaces become $N(1/g^2N)^{S_1+S_2}$ for $N\geq 3$, and $2N(1/g^2N)^{S_1+S_2}$ for $N=2$, respectively.
The difference of factor $2$ in front of $N$ for $N=2$ arises from the non-oriented nature of the plaquettes  to conclude the $N=2$ result:
\begin{align}
   4\left(   \frac{1}{2g^2} \right)^{S_1+S_2} .
\end{align}

\begin{figure}[tbp]
\begin{center}
\includegraphics[height=4.0cm]{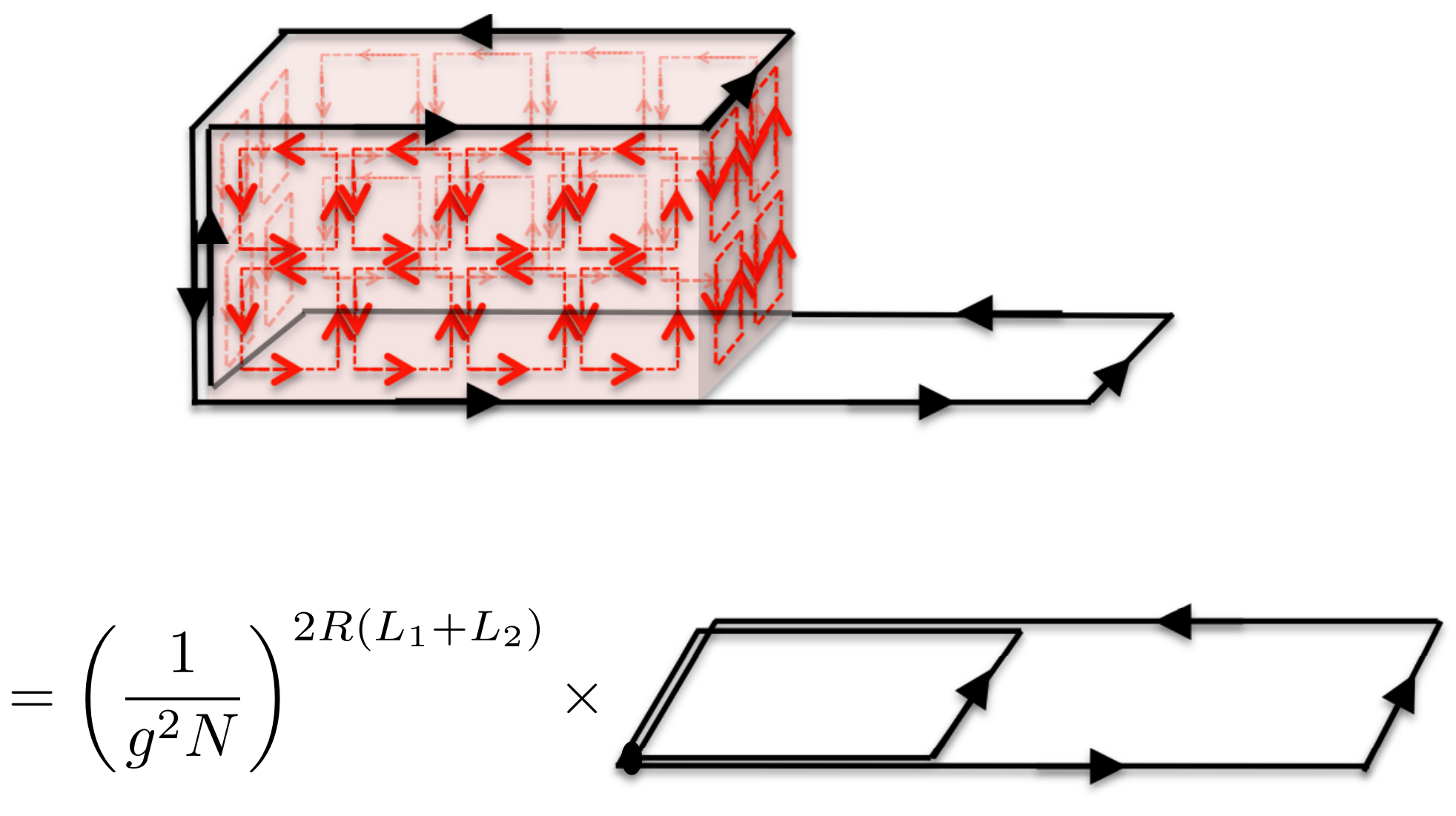}
\caption{  
Another diagram which contributes to a shifted double-winding Wilson loop average $\langle W(C_1\times C_2 ) \rangle_{R \ne 0}$  in the strong coupling expansion of the $SU(N)$ lattice Yang-Mills theory.
}
\label{Dw-fig9}
\end{center}
\end{figure}
Another type of diagram which also gives a leading contribution in the strong coupling expansion is given by a set of plaquettes tiling the minimal surface $S_1-S_2$ and the four sides with the area $2R(L_1+L_2)$ of a cuboid with a height $R$, whose bottom is a rectangular of size  $L_1\times L_2$, as shown in the upper panel of Fig.\ref{Dw-fig9}.
After group integrations for the links on the side surfaces giving a factor $(1/g^2N)^{2R(L_1+L_2)}$, this diagram is equivalent to a coplanar double-winding Wilson loop, as shown in the lower panel of Fig.\ref{Dw-fig9}.
The expectation value of this type of a coplanar double-winding Wilson loop is already calculated in the previous subsection, and the results are eq.(\ref{sce-su2r-1}) for $SU(2)$, eq.(\ref{sce-su3-1}) for $SU(3)$, and eq.(\ref{sce-su4-1}) for $SU(4)$, respectively. Consequently, the diagram of Fig.\ref{Dw-fig9} yields the contribution for $N=2$:
\begin{align}
\left(  \frac{1}{2g^2} \right)^{2R(L_1+L_2)} 
\left\{
2p_2 \left(  \frac{1}{2g^2} \right)^{S_1-S_2}
+2q_2 \left(  \frac{1}{2g^2} \right)^{S_1+S_2} 
\right\} .
\label{sce-su2-z1}
\end{align}

\begin{figure}[tbp]
\begin{center}
\includegraphics[height=6.0cm]{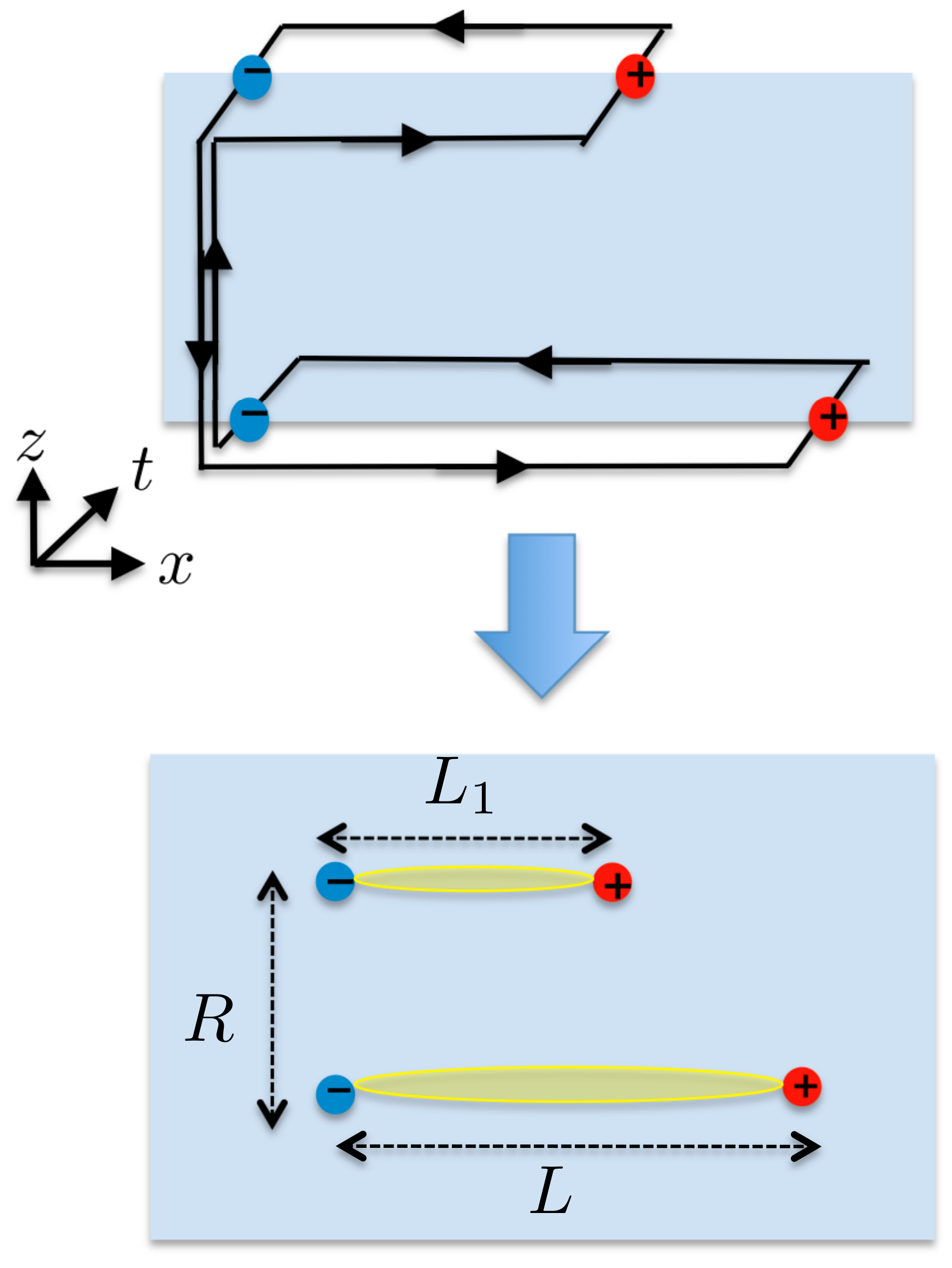}
\caption{ 
A shifted double-winding Wilson loop as a probe for interactions between two flux tubes.  
}
\label{Dw-fig11}
\end{center}
\end{figure}

To summarize the above discussion, the expectation value of the shifted double-winding loop $\langle W(C_1\times C_2 ) \rangle_{R \ne 0}$
from diagrams as shown in Fig.\ref{Dw-fig10} and Fig.\ref{Dw-fig9} becomes for $N=2$, 
\begin{align}
SU(2): \ 
&\langle W(C_1\times C_2 ) \rangle_{R\ne 0} 
\nonumber\\  
=&  4\left(   \frac{1}{2g^2} \right)^{S_1+S_2}
+2p_2\left(   \frac{1}{2g^2} \right)^{S_1-S_2+2R(L_1+L_2)}
\nonumber\\&
+2q_2\left(   \frac{1}{2g^2} \right)^{S_1+S_2+2R(L_1+L_2)} + \cdots.
\label{sce-su2-z1b}
\end{align}
Note that the $R \to 0$ limit of eq.(\ref{sce-su2-z1b}) does not agree with the coplanar result eq.(\ref{sce-su2r-1}), 
although the sum of the second and third terms in eq.(\ref{sce-su2-z1b}) from the diagram of Fig.\ref{Dw-fig9} reproduce the coplanar result eq.(\ref{sce-su2r-1}) in the limit $R \to 0$.
This is because the first term in eq.(\ref{sce-su2-z1b}) coming from the diagram of Fig.\ref{Dw-fig10} does not have in the limit $R \to 0$ the counterpart of the strong coupling expansion in the coplanar case and hence contributes only to the shifted case with $R \ne 0$. 

For $SU(2)$ gauge group, especially, we perform the detailed study on the $R$-dependence of a shifted double-winding Wilson loop average $\langle W(C_1\times C_2 ) \rangle_{R \ne 0} $. 
In what follows, we rewrite $L_2$  
into $T$,
\begin{align}
 T:=L_2 .
\end{align}
Let us imagine $T$ direction be time $t$-axis, $L$ and $L_1$ direction be spatial $x$-axis, and 
$R$ direction be also space $z$-axis as seen in top side in Fig.\ref{Dw-fig11}. 
As is explained in \cite{GH15}, the shifted double-winding Wilson loop at a fixed time can be interpreted as a tetra-quark system  consisting of two static quarks and two static antiquarks.
The pairs of quark-antiquarks are connected by a pair of color flux tubes, as seen in the bottom side in Fig.\ref{Dw-fig11}. 
We study how interactions between the two color flux tubes change, when the distance $R$ is varied.

\begin{figure}[tbp]
\begin{center}
\includegraphics[height=3.0cm]{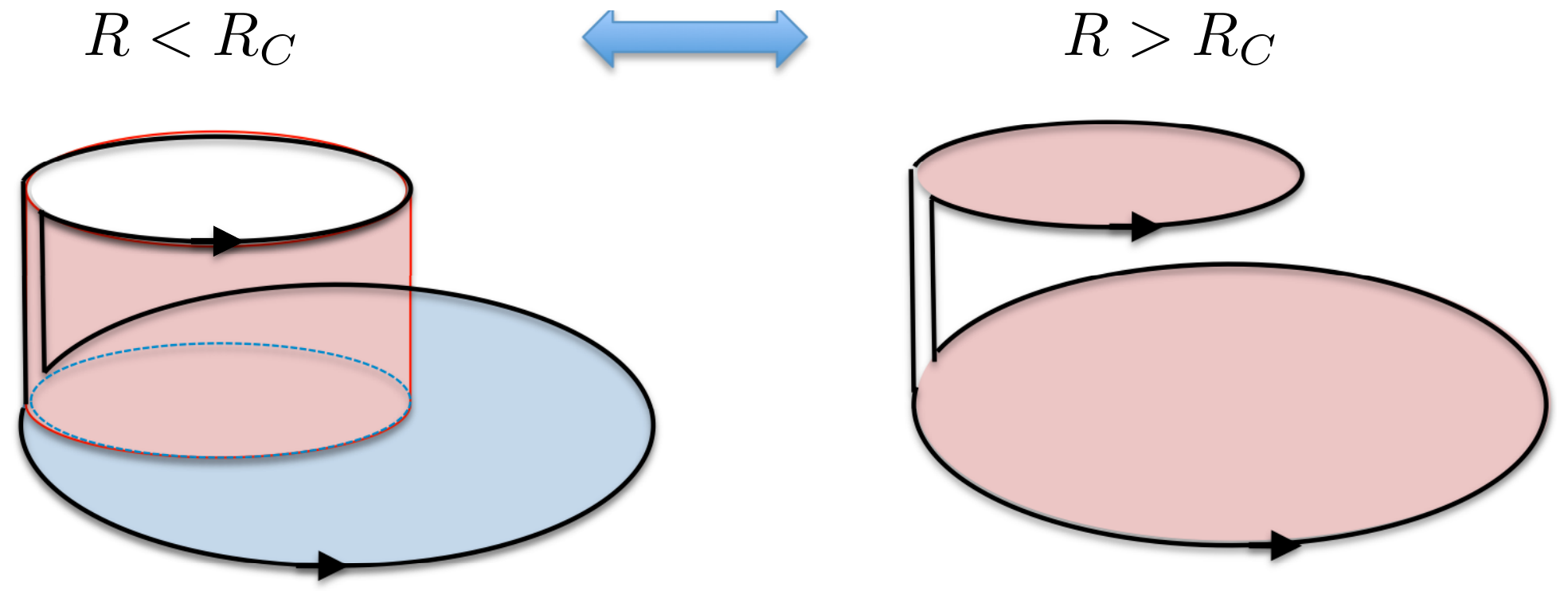}
\caption{ 
Lowest order diagrams giving the dominant contribution to  a shifted double-winding Wilson loop average $\langle W(C_1\times C_2 ) \rangle_{R \ne 0} $.
The dominant diagram switches at a certain value $R_c$ of $R$ from left to right.
}
\label{Dw-fig12}
\end{center}
\end{figure}

\begin{figure}[tbp]
\begin{center}
\includegraphics[height=5.0cm]{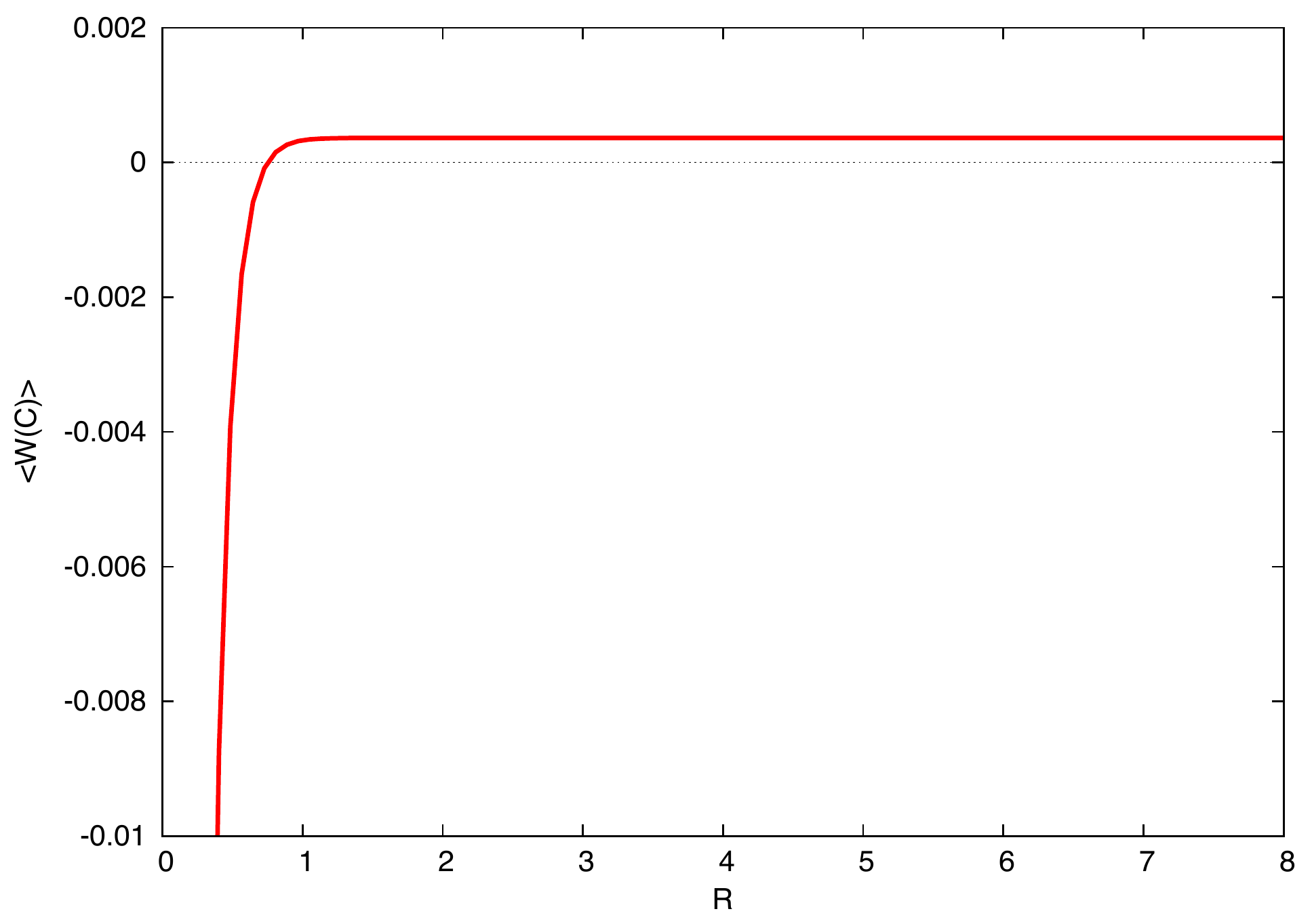}
\caption{ 
$R$-dependence of of a shifted double-winding Wilson loop average  $\langle W(C_1\times C_2 ) \rangle$ in the $SU(2)$ lattice gauge theory obtained from the strong coupling expansion for $1/2g^2=2.5/8$, $L=5$, $L_2=1$ and $L_1=3$.
}
\label{Dw-fig13}
\end{center}
\end{figure}

We find that 
the second term in eq.(\ref{sce-su2-z1b}) dominates for 
$R<R_C := \frac{L_1}{1+L_1/T}$, 
and the first term in eq.(\ref{sce-su2-z1b}) dominates for $R>R_C$,
because the comparison of the two exponents of these terms for $S_1=LT$ and $S_2=L_1 T$ reads 
\begin{align}
&  S_1-S_2+2R(L_1+L_2) < S_1+S_2
\Longrightarrow 
  R(L_1+L_2) <  S_2
 \nonumber\\&
\Longrightarrow 
     R (L_1+T) < L_1 T 
 \Longrightarrow 
   R < \frac{L_1}{1+L_1/T} := R_C ,    
\end{align}
where we have neglected the third (higher order) term in eq.(\ref{sce-su2-z1b}) for the naive estimate of $R_C$.
This means that the left diagram of Fig.\ref{Dw-fig12}   dominates for $R<R_C$, and the right diagram of Fig.\ref{Dw-fig12}  dominates for $R>R_C$.
Therefore, the dominant diagram switches from left to right  at a certain value $R_C$ of $R$ as $R$ increases, just like the minimal surface spanned by a soap film.

In Fig.\ref{Dw-fig13}, we plot the $R$-dependence eq.(\ref{sce-su2-z1b}) of a shifted double-winding Wilson loop average $\langle W(C_1\times C_2 ) \rangle$ for fixed $L$, $L_1$, and $L_2$ in the $SU(2)$ lattice gauge theory.
The  second and third terms in eq.(\ref{sce-su2-z1b})  have $R$-dependence, but the first term  in eq.(\ref{sce-su2-z1b})
does not depend on $R$. 
Therefore, the plot gets flattened for $R\geq R_C \sim 1$, 
which is consistent with Fig.\ref{Dw-fig12}.
This behavior does not depend on the number of color $N$.
In fact, $SU(3)$ and $SU(4)$ cases are given as follows.

\begin{align}
SU(3): \ 
& \langle W(C_1\times C_2 ) \rangle_{R \ne 0} 
\nonumber\\  
=& 
3\left(   \frac{1}{3g^2} \right)^{S_1+S_2}
+p_3 \left(  \frac{1}{3g^2} \right)^{S_1+2R(L_1+L_2)}
\nonumber\\&
+q_3 \left(  \frac{1}{3g^2} \right)^{S_1+S_2+2R(L_1+L_2)} + \cdots,
\label{sce-su3-z1}
\end{align}

\begin{align}
SU(4): \ 
&\langle W(C_1\times C_2 ) \rangle_{R \ne 0} 
\nonumber\\  
=&
4\left(   \frac{1}{4g^2} \right)^{S_1+S_2}
+p_4 \left(  \frac{1}{4g^2} \right)^{S_1+S_2+2R(L_1+L_2)}
\nonumber\\&
+q_4 \left(  \frac{1}{4g^2} \right)^{S_1+S_2+2R(L_1+L_2)} + \cdots.
\label{sce-su4-z1}
\end{align}

\begin{figure}[tbp]
\begin{center}
\includegraphics[height=5.0cm]{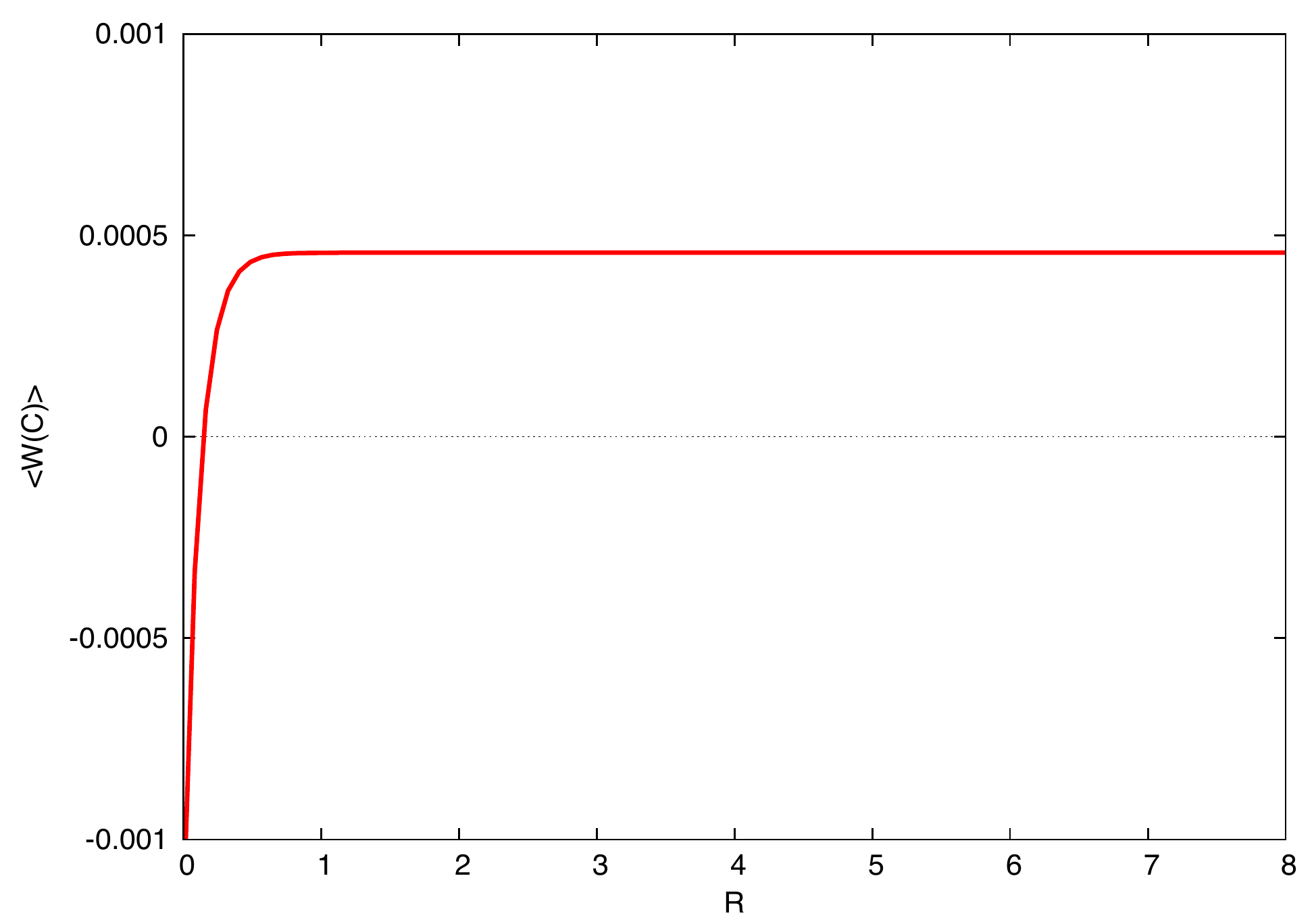}
\caption{\small 
$R$-dependence of of a shifted double-winding Wilson loop average  $\langle W(C_1\times C_2 ) \rangle$ in the $SU(3)$ lattice gauge theory obtained from the strong coupling expansion for $1/3g^2=6.0/18$, $L=5$, $L_2=1$ and $L_1=3$. 
}
\label{Dw-fig13-su3}
\end{center}
\end{figure}

In Fig.\ref{Dw-fig13-su3}, we also plot the $R$-dependence eq.(\ref{sce-su3-z1}) of a shifted double-winding Wilson loop average $\langle W(C_1\times C_2 ) \rangle$ for fixed $L$, $L_1$, and $L_2$ in the $SU(3)$ lattice gauge theory.

In general,  $\langle W(C_1\times C_2 ) \rangle_{R \ne 0}$ for $N \geq 3$ becomes

\begin{align}
&\langle W(C_1\times C_2 ) \rangle_{R \ne 0} 
\nonumber\\  
=&
N\left(   \frac{1}{g^2N} \right)^{S_1+S_2}
+\left(  \frac{1}{g^2N} \right)^{2R(L_1+L_2)} \times
\nonumber\\&
\left\{
p_N \left(  \frac{1}{g^2N} \right)^{(N-2)S_2+S_1-S_2}
+q_N \left(  \frac{1}{g^2N} \right)^{S_1+S_2}
\right\} + \cdots.
\label{sce-suN-z1}
\end{align}

\subsection{Numerical simulation}

\begin{figure}[tbp]
\begin{center}
\includegraphics[height=5.0cm]{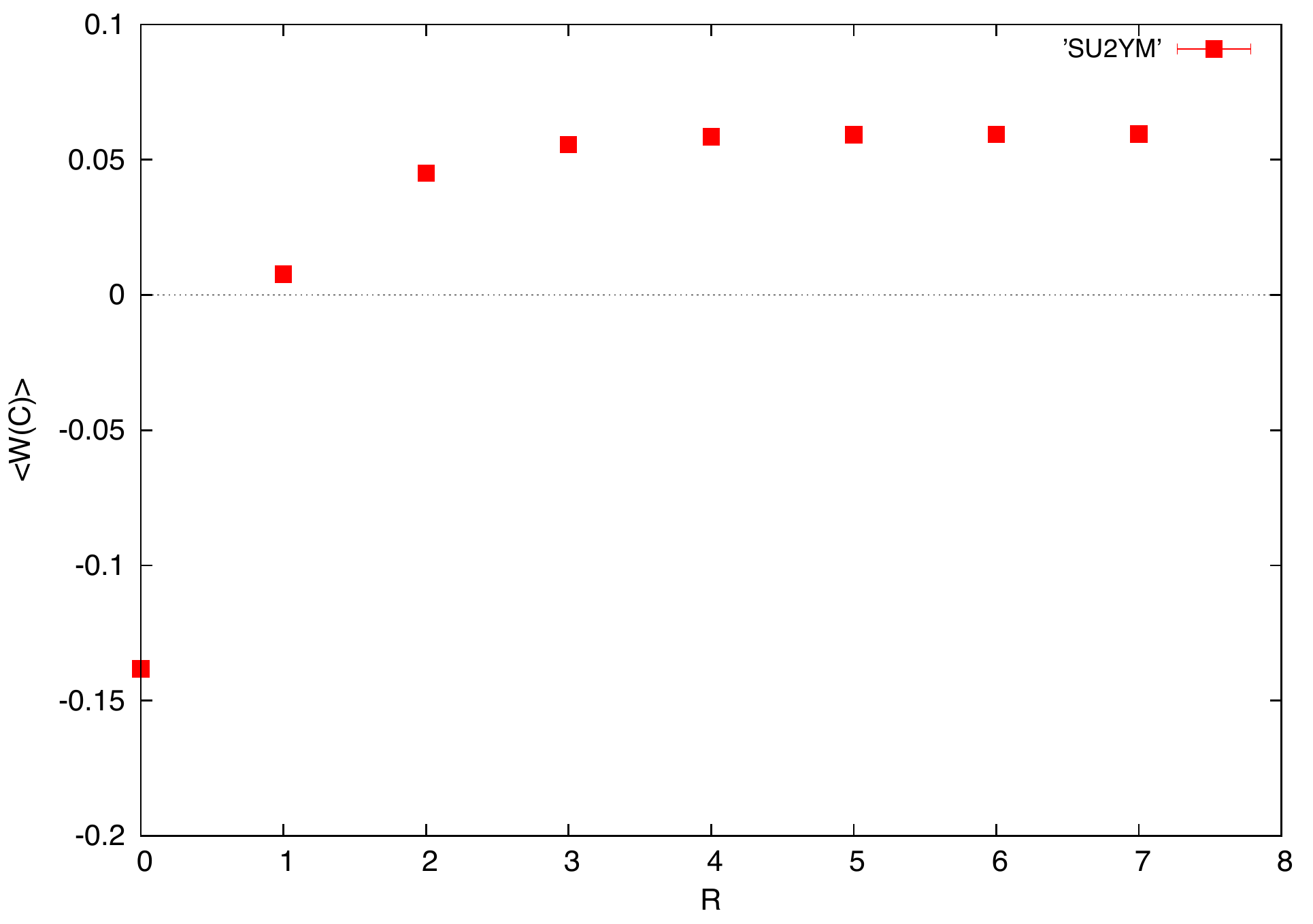}
\caption{ 
$R$-dependence of a shifted double-winding Wilson loop average $\langle W(C_1\times C_2 ) \rangle$  in the $SU(2)$ lattice gauge theory obtained from numerical simulations on a lattice of size $24^4$ at  $\beta=2.5$ for fixed $L=5$, $T(=L_2)=2$, and $L_1=3$.
}
\label{numerical-fig5}
\end{center}
\end{figure}
\begin{figure}[tbp]
\begin{center}
\includegraphics[height=5.0cm]{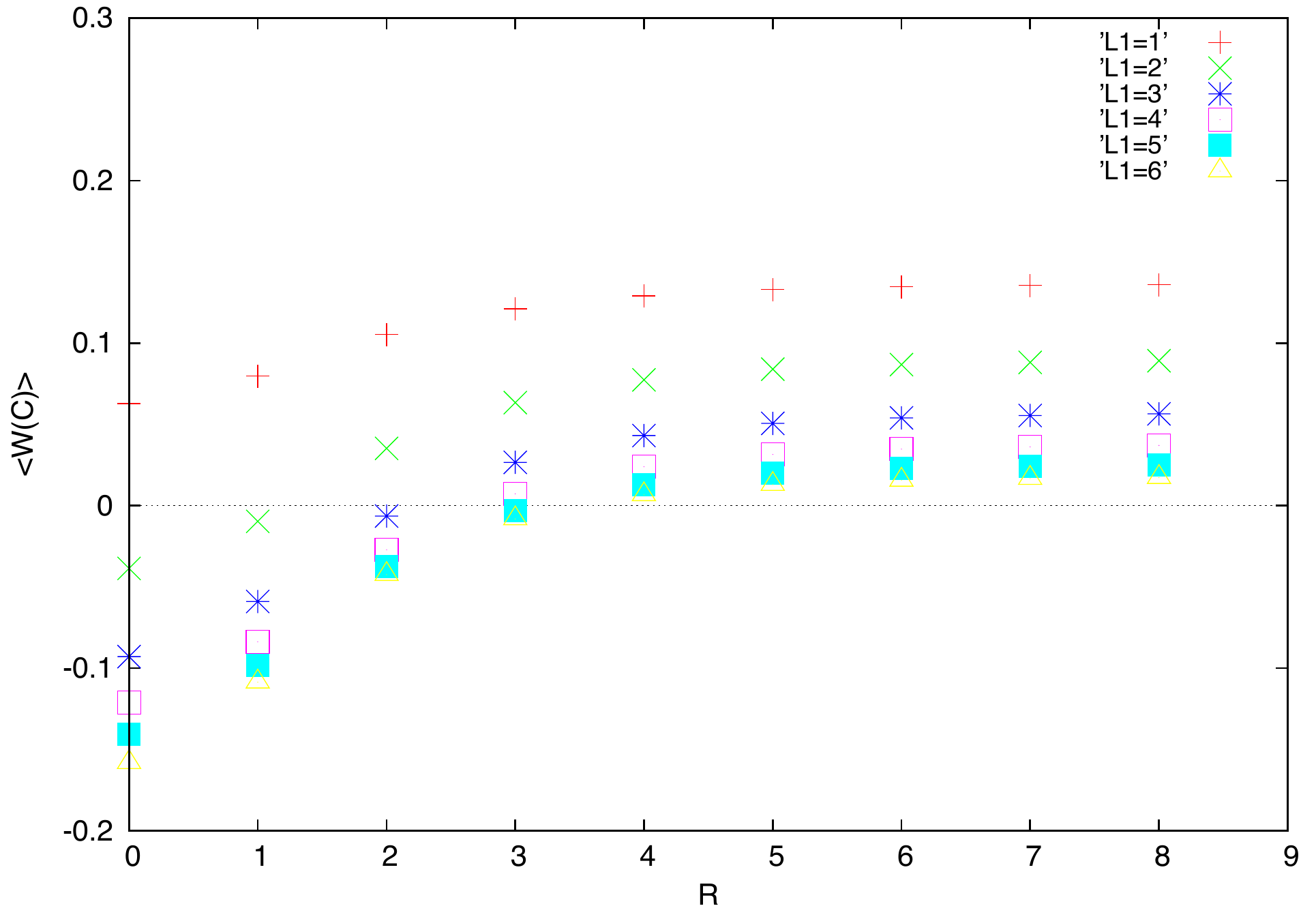}
\caption{ 
$R$-dependence of a shifted double-winding Wilson loop average $\langle W(C_1\times C_2 ) \rangle$ in the $SU(3)$ lattice gauge theory obtained from numerical simulations on a lattice of size $24^4$ at  $\beta=6.2$ $L=8$, $T(=L_2)=8$, and $L_1=1\sim 6$ from top to bottom.
}
\label{numerical-fig6}
\end{center}
\end{figure}

Next, we examine the $R$-dependence of $\langle W(C_1\times C_2 ) \rangle$ based on numerical simulations on a lattice.

\underline{$SU(2)$}:  
In order to calculate the shifted double-winding Wilson loop average, we use the same gauge field configurations as those used in calculating the coplanar double-winding Wilson loop. 
However, we have used APE smearing method ($N=5$, $\alpha=0.1$) as a noise reduction technique.
Fig.\ref{numerical-fig5} gives the plots obtained for the $\langle W(C_1\times C_2 ) \rangle$ for various values of $R$ 
where we have fixed $L=5$, $T(=L_2)=2$, $L_1=3$.
We see that the behavior of  data in Fig.\ref{numerical-fig5} is consistent with the analytical result given in Fig.\ref{Dw-fig13}.

\underline{$SU(3)$}:  
Similarly, Fig.\ref{numerical-fig6} shows the obtained plot for the $\langle W(C_1\times C_2 ) \rangle$ for various value of $R$ for 
$SU(3)$ case,
when we choose parameters $L=8$, $T(=L_2)=8$, $L_1=1\sim6$.
We see that the data in Fig.\ref{numerical-fig6} also consistent with the analytical result given in Fig.\ref{Dw-fig13-su3} for sufficiently large  areas $S_1$ and $S_2$.

\section{Conclusion and Discussion}

In this paper, we have studied the  double-winding Wilson loops in $SU(N)$ lattice Yang-Mills gauge theory by using both strong coupling expansion and numerical simulation.

First of all, we have examined how the area law falloff of a ``coplanar'' double-winding Wilson loop average depends on the number of color $N$, by changing the size of minimal area $S_2$ of loop $C_2$.
We have reconfirmed the difference-of-areas law for $N=2$, and have found new results that  ``max-of-areas law''  for $N=3$ and sum-of-areas law for $N\geq 4$.

Moreover, we have considered a ``shifted''  double-winding Wilson loop, where two contours  are displaced from one another in a transverse direction.
We have evaluated its average by changing the distance of a transverse direction, and have found that their long distance behavior doesn't depend on  the number of color $N$, but the short distance behavior depends on $N$.

It should be remarked that this ``shifted''  double-winding Wilson loop may contain an information about interactions between two color flux tubes. For this purpose, we need to accumulate more data on the fine lattices with more larger size. 

Originally, one of reasons why Greensite and H\"ollwieser  considered the double-winding Wilson loops seems to be that they want to evaluate monopole confinement mechanism in lattice $SU(2)$ gauge theory.
They have considered an operator which simply replaces $SU(2)$ link variable $U_{n,\mu}$ with the Abelian variable $u_{n,\mu}$ as an ``Abelian'' double-winding Wilson loop, and have shown that the expectation value of such a naive operator obeys the sum-of-areas law.
But, it is known that such naive operator should work only for a single-winding Wilson loop in the fundamental representation.
Recently, Matsudo and his collaborators \cite{Matsudo2019} have given the explicit expression for the Abelian operator which reproduces the full Wilson loop average in higher representations, which is suggested by the gauge-covariant field decomposition and the non-Abelian Stokes theorem (NAST) for the Wilson loop operator.
Similarly, we hope that a correct form of the Abelian operator for a double-winding Wilson loop  can be found in the similar way.
When we change the line integral to the surface integral,  our considerations of the diagrams which give the leading contribution to the strong coupling expansion seems to be useful to construct the NAST for a double-winding Wilson loop.
These results will be discussed in a forthcoming paper.

\section*{Acknowledgments}

This work was  supported by Grant-in-Aid for Scientific Research, JSPS KAKENHI Grant Number (C) No.19K03840 and No.15K05042. 

\appendix
\renewcommand{\theequation}{{\thesection.\arabic{equation}}}
\renewcommand{\thetable}{{\thesection.\arabic{table}}}
\renewcommand{\thefigure}{{\thesection.\arabic{figure}}}

\setcounter{figure}{0}

\begin{widetext}

\section{$SU(N)$ group integrals and useful formulae}
\crefname{equation}{}{}
\crefrangelabelformat{equation}{(#3#1#4--#5#2#6)}
\crefname{figure}{Fig.}{Figs.}

In order to perform the strong coupling expansion in the lattice gauge theory, we must calculate the following   integrations for the polynomials of group matrix elements over each links:
\begin{align}
I=\int dU \ U_{i_1j_1} \cdots U_{i_nj_n}(U^{-1})_{k_1l_1} \cdots (U^{-1})_{k_ml_m},
\label{chap1-1}
\end{align}
where  $U_{ij}$ ($i,j=1,2, \cdots, N$) denotes a matrix element of a matrix $U \in SU(N)$ belonging to the $SU(N)$ group with the property $U^{-1}=U^{\dagger}$, 
and $dU$ is an invariant measure (Haar measure) on the compact group which is left-invariant
\begin{align}
\int dU \ f(U) =\int dU \ f(VU)   \quad (^{\forall}V \in SU(N)),
\label{chap1-2}
\end{align}
and right-invariant
\begin{align}
\int dU \ f(U) =\int dU \ f(UV)   \quad (^{\forall}V \in SU(N)) .
\label{chap1-3}
\end{align}
We can normalize the measure such that
\begin{align}
\int dU =1.
\label{chap1-4}
\end{align}

By using properties of the invariant measure, Creutz has shown that eq.(\ref{chap1-1}) can be evaluated by the 
following formula \cite{Creutz:text,Creutz:1978}:
\begin{align}
I  =  (\partial_{j_1i_1}\cdots \partial_{j_ni_n}\cdot {\rm cof}(\partial)_{l_1k_1}\cdots {\rm cof}(\partial)_{l_mk_m})
   \sum_{i=0}^{\infty}\frac{2!3!\cdots (N-1)!}{i!(i+1)!\cdots (i+N-1)!}|J|^i\vert_{J=0} ,
\label{chap2-4}
\end{align}
where $J$ is a source variable and is an arbitrary $N \times N$ matrix, $|J|=\det(J)$, 
$\partial_{ji}\equiv \partial / \partial J_{ji}$, and ${\rm cof}(\partial)$ is a cofactor of $\partial$, 
respectively.

We list some of explicit results from the above formula as 
\begin{align}
& \int dU \ 1 =1,
\label{A-sub2-5-1}  
\\
& \int dU \ U_{ab} =0,
\label{A-sub2-5-2}
\\
& \int dU \ U_{ab}U^{\dagger}_{kl} =\frac{1}{N}\delta_{al}\delta_{bk},
\label{A-sub2-5-3} 
\\
& \int dU \ U_{a_1b_1}U_{a_2b_2}\cdots U_{a_Nb_N} =\frac{1}{N!}\epsilon_{a_1a_2\cdots a_N}\epsilon_{b_1b_2\cdots b_N},
\label{A-sub2-5-4}
\\
& \int dU \ U_{a_1b_1}U_{a_2b_2}\cdots U_{a_Mb_M} =0,  \quad M\ne 0 \ ({\rm mod}N),
\label{A-sub2-5-5}
\\
& \int dU \ U_{ab}U_{cd}U^{\dagger}_{ij}U^{\dagger}_{kl} =
\frac{1}{(N^2-1)}\left[
\delta_{aj}\delta_{bi}\delta_{cl}\delta_{dk}+\delta_{al}\delta_{bk}\delta_{cj}\delta_{di}
-\frac{1}{N}(
\delta_{aj}\delta_{bk}\delta_{cl}\delta_{di}+\delta_{al}\delta_{bi}\delta_{cj}\delta_{dk}
)\right] .
\label{A-sub2-5-6}
\end{align}
The last eq.(\ref{A-sub2-5-6}) consist for $N>2$. For $N=2$, 

\begin{align}
\int dU \ U_{ab}U_{cd}U^{\dagger}_{ij}U^{\dagger}_{kl}
=&
\frac{1}{(N^2-1)} \left[
\delta_{aj}\delta_{bi}\delta_{cl}\delta_{dk}+\delta_{al}\delta_{bk}\delta_{cj}\delta_{di}
-\frac{1}{N}(
\delta_{aj}\delta_{bk}\delta_{cl}\delta_{di}+\delta_{al}\delta_{bi}\delta_{cj}\delta_{dk}
)\right]   
+\left(  \frac{1}{N!}\right)^{2}\epsilon_{ac}\epsilon_{bd}\epsilon_{ik}\epsilon_{jl}.
\label{A-sub2-5-6b}
\end{align}

Following relation can be shown by using property of invariant measure,
\begin{align}
\int dU \ f(U^{-1}) = \int dU \ f(U).
\label{A-sub2-3-3}
\end{align}
From this relation, we also obtain,
\begin{align}
& \int dU \ U^{\dagger}_{ab} =0,
\label{A-sub2-5-2b} \\
& \int dU \ U^{\dagger}_{a_1b_1}U^{\dagger}_{a_2b_2}\cdots U^{\dagger}_{a_Nb_N} 
=\frac{1}{N!}\epsilon_{a_1a_2\cdots a_N}\epsilon_{b_1b_2\cdots b_N}.
\label{A-sub2-5-4b}
\end{align}

The following more practical formulae are useful to calculate the expectation value of 
double-winding Wilson loop by using strong coupling expansion.
Let  $X,Y,A,B$ be elements of $SU(N)$ group.
From eq.(\ref{A-sub2-5-4}), we find
\begin{align}
& \int dU {\rm tr}(  XUYU) 
=X_{ab}Y_{cd}\int dU \ U_{bc}U_{da}
=\delta_{N,2} \frac{1}{N}\epsilon_{ca}X_{ab}Y_{cd}\epsilon_{bd} ,
\label{eq:BP0}\\
& \int dU {\rm tr}(  XU) {\rm tr}(YU)   
=X_{ab}Y_{cd}\int dU \ U_{ba}U_{dc}
=\delta_{N,2} \frac{1}{N}  \epsilon_{ac}\epsilon_{bd}X_{ab}Y_{cd}.
\label{eq:BP0a}
\end{align}
From eq.(\ref{A-sub2-5-3}), we find
\begin{align}
\int dU {\rm tr}( XU) {\rm tr}(YU^{\dagger})
=X_{ab}Y_{lk}\int dU \ U_{ba}U_{kl}^{\dagger}
=X_{ab}Y_{lk}\frac{1}{N}\delta_{bl}\delta_{ak}
=\frac{1}{N}{\rm tr}(XY).
\label{eq:BP1}
\end{align}

From eq.(\ref{A-sub2-5-6}),  we find for $N>2$,
\begin{align}
&  \int dU {\rm tr}(AU) {\rm tr}(BU){\rm tr}(XU^{\dagger})  {\rm tr}( YU^{\dagger}) 
\nonumber\\
&  =A_{ab}B_{cd}X_{ij}Y_{kl}\int dU \ U_{ba}U_{dc}U^{\dagger}_{ji}U^{\dagger}_{lk}
\nonumber\\
&  =A_{ab}B_{cd}X_{ij}Y_{kl}\frac{1}{N^{2}-1}
\left[  \delta_{bi}\delta_{aj}\delta_{dk}\delta_{cl}+\delta_{bk}\delta_{al}\delta_{di}\delta_{cj}
-\frac{1}{N}\left(  \delta_{bi}\delta_{al}\delta_{dk}\delta_{cj}+\delta_{bk}\delta_{aj}\delta_{di}\delta_{cl}\right)  
\right] \nonumber\\
&  =\frac{1}{N^{2}-1}
\left[  {\rm tr}(AX) {\rm tr}(BY)  +{\rm tr}(AY) {\rm tr}( BX)  -\frac{1}{N}\left(  {\rm tr}(AXBY)  +{\rm tr}(AYBX)\right)  \right] , 
\label{eq:BP2}
\end{align}
\begin{align}
&  \int dU {\rm tr}(AUBU){\rm tr}( XU^{\dagger})  {\rm tr}( YU^{\dagger})
 \nonumber\\
&  =A_{ab}B_{cd}X_{ij}Y_{kl}\int dU \ U_{bc}U_{da}U^{\dagger}_{ji}U^{\dagger}_{lk}
\nonumber\\
&  =A_{ab}B_{cd}X_{ij}Y_{kl}\frac{1}{N^{2}-1}
\left[  
\delta_{bi}\delta_{cj}\delta_{dk}\delta_{al}+\delta_{bk}\delta_{cl}\delta_{di}\delta_{aj}
-\frac{1}{N}\left(  \delta_{bi}\delta_{cl}\delta_{dk}\delta_{aj}+\delta_{bk}\delta_{cj}\delta_{di}\delta_{al}\right)  
\right] \nonumber\\
&  =\frac{1}{N^{2}-1}
\left[  
{\rm tr}(AXBY)  +{\rm tr}(AYBX)  -\frac{1}{N}(  {\rm tr}( AX) {\rm tr}(BY)+{\rm tr}( AY) {\rm tr}(BX))  
\right],
\label{eq:BP3}
\end{align}
\begin{align}
&  \int dU {\rm tr}(AUBU) {\rm tr}( XU^{\dagger}YU^{\dagger})
\nonumber\\
&  =A_{ab}B_{cd}X_{ij}Y_{kl}\int dU \ U_{bc}U_{da}U^{\dagger}_{jk}U^{\dagger}_{li}
\nonumber\\
&  =A_{ab}B_{cd}X_{ij}Y_{kl}\frac{1}{N^{2}-1}
\left[  
\delta_{bk}\delta_{cj}\delta_{di}\delta_{al}+\delta_{bi}\delta_{cl}\delta_{dk}\delta_{aj}
-\frac{1}{N}
\left(  
\delta_{bk}\delta_{cl}\delta_{di}\delta_{aj}+\delta_{bi}\delta_{cj}\delta_{dk}\delta_{al}
\right)  \right] \nonumber\\
&  =\frac{1}{N^{2}-1}
\left[ 
{\rm tr}(AY){\rm tr}(BX)+{\rm tr}(AX){\rm tr}(BY)-\frac{1}{N}\left(  {\rm tr}( AYBX)+{\rm tr}( AYBX)  \right) 
\right] . 
\label{eq:BP4}%
\end{align}


\section{Explicit calculation of the coefficient $q_N$}
\crefname{equation}{}{}
\crefrangelabelformat{equation}{(#3#1#4--#5#2#6)}
\crefname{figure}{Fig.}{Figs.}

In this section, we show explicitly how eq.(\ref{sce-q_N}) is obtained. 
From eq.(\ref{chap3-3}) and eq.(\ref{chap3-5-2}), a contribution to a coplanar double-winding Wilson loop average 
$\langle W(C_1\times C_2 ) \rangle$ from the bottom panel of Fig.\ref{Dw-fig7} is expressed as 
\begin{align}
 \langle W(C_1\times C_2 ) \rangle_{q_N} = \int\prod_{\ell \in S_1}dU_{\ell} \ W(C_1\times C_2 ) \cdot
 \prod_{p_j\in (S_1-S_2)} \left[  \frac{1}{g^2} {\rm tr} (U^{\dagger}_{p_j}) \right] \cdot
 \prod_{p_k\in S_2} \left\{  \frac{1}{2!} \left[ \frac{1}{g^2} {\rm tr} (U^{\dagger}_{p_k})\right]^2  \right\},
\label{ap-B-1}
\end{align}
where $U^{\dagger}_{p_j}$ and $U^{\dagger}_{p_k}$ denote  respectively plaquette variables on $(S_1-S_2)$ and $S_2$ areas.
Here note that $U_{p}^\dagger$ represents the plaquette variable for the plaquette ${p}$ with the clockwise orientation. 

First, integration with respect to the link variables $\{ U_\ell \}$ on the $(S_1-S_2)$ area can be performed with the same 
technique of the strong coupling expansion as that for the fundamental Wilson loop to obtain
\begin{align}
 \langle W(C_1\times C_2 ) \rangle_{q_N} = 
 \left(  \frac{1}{g^2N} \right)^{S_1-S_2}
  \langle W(C_2\times C_2 ) \rangle_{q_N} ,
\label{ap-B-2}
\end{align}
where we have defined 
\begin{align}
 \langle W(C_2\times C_2 ) \rangle_{q_N} := \int\prod_{\ell \in S_2}dU_{\ell} \ W(C_2\times C_2 ) \cdot
 \prod_{p_k\in S_2} \left\{  \frac{1}{2!} \left[ \frac{1}{g^2} {\rm tr} (U^{\dagger}_{p_k})\right]^2  \right\}.
\label{ap-B-3}
\end{align}

\begin{figure}[tbp]
\begin{center}
\includegraphics[height=3.0cm]{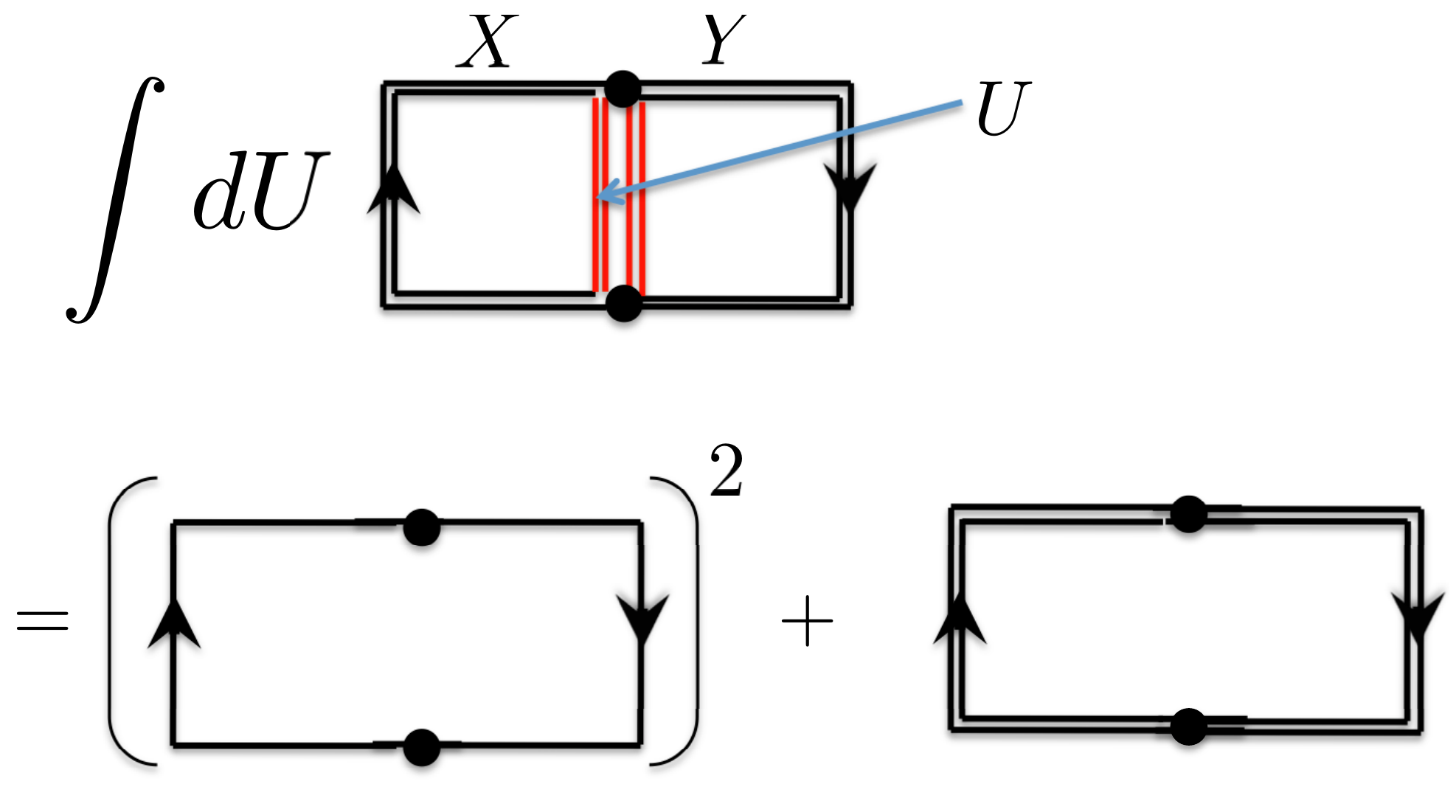}
\caption{ 
Diagrammatic representation of the integration rule $\tilde{W}_2$ eq.(\ref{ap-B-4}) for the product of two double-plaquettes with the same clockwise orientation: 
Integration is performed over the link variables $U$ on the link which is common to two double-plaquettes with the same  clockwise orientation. 
By decomposing the path-ordered product of the link variables along the loop, 
the plaquette variables for the single plaquette $p_1$ and $p_2$ to the left and right of $U$ is respectively represented by 
${\rm tr}(U_{p_1}^{\dagger}):={\rm tr}(U^{\dagger}X)$ and ${\rm tr}(U_{p_2}^{\dagger}):={\rm tr}(YU)$.
Here $X$ and $Y$ represent the products of the link variables along staple-shaped paths with the same orientations.
}
\label{appendixB-fig1}
\end{center}
\end{figure}

Next, we perform the integration in eq.(\ref{ap-B-3}) over the link variables $\{ U_\ell\}$ inside of the $S_2$ area, which excludes the links on the loop $C_2=\partial S_2$ (the boundary of $S_2$).
As shown in Fig.\ref{appendixB-fig1}, performing the integration with respect to the link variables $U$ on the link 
which is common to two double-plaquettes with the same clockwise orientation  using eq.(\ref{eq:BP2}), 
 we obtain
\begin{align}
\tilde{W}_2
&:= \int dU \left\{{\rm tr}(U^{\dagger}X)\right\}^2 \cdot  \left\{{\rm tr}(YU)\right\}^2 
= \alpha_2 \left\{  {\rm tr}(YX) \right\}^2 + \beta_2 {\rm tr} (YXYX) 
:= \alpha_2 W(D_2)^2 + \beta_2 W(D_2\times D_2),
\label{ap-B-4} 
\end{align}
where 
\begin{align}
\alpha_2 = \frac{2}{N^2-1} , \quad \beta_2 =-\frac{2}{N(N^2-1)}.
\label{ap-B-5}  
\end{align}
Here $D_2$ represents the loop as the boundary of 
a $2\times 1$ rectangle obtained by combining two fundamental (square) plaquettes which are adjacent to the link $U$.
Then $W(D_2)$ and $W(D_2\times D_2)$  respectively stand for the single-winding Wilson 
loop and double-winding Wilson loop along the loop $D_2$ where the Wilson loop means the trace of the product of link variables on the relevant loop.

\begin{figure}[tbp]
\begin{center}
\includegraphics[height=3.0cm]{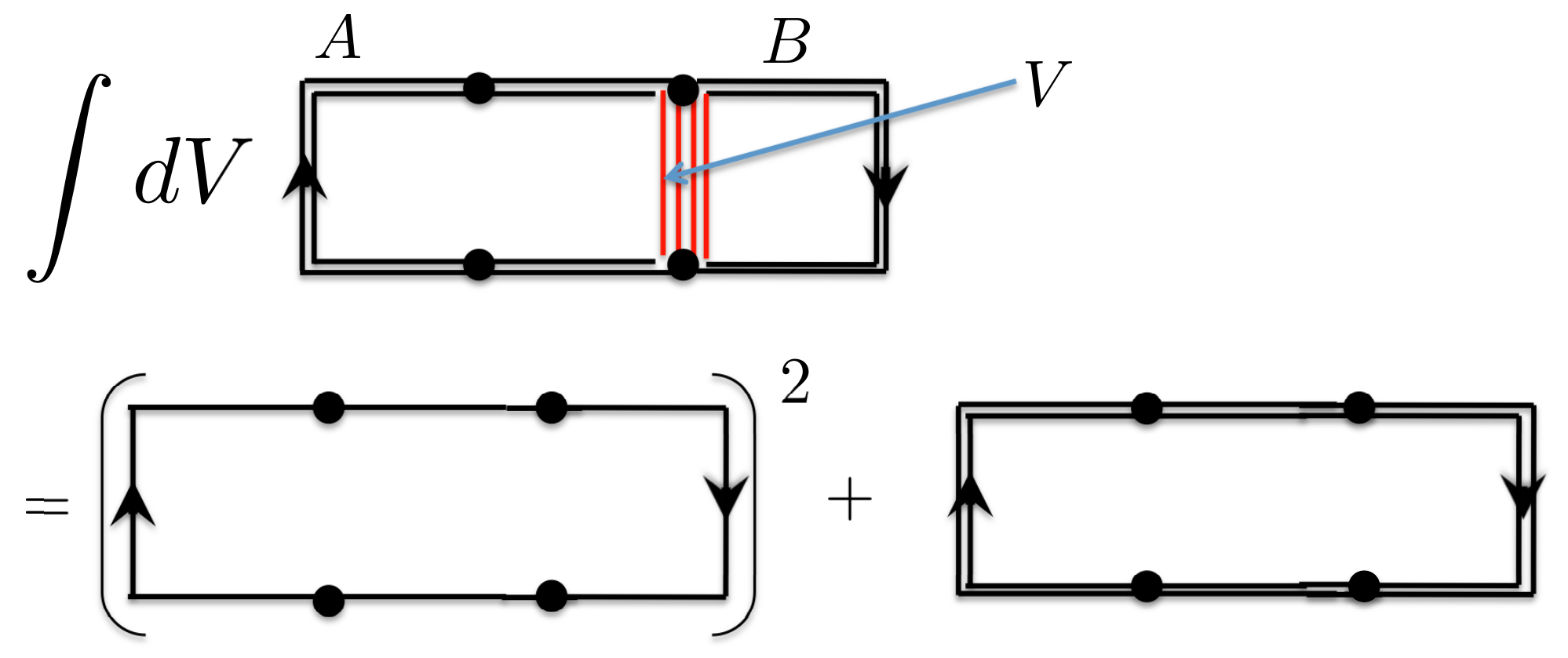}
\caption{ 
Diagrammatic representation of the integration rule eq.(\ref{ap-B-6}) for the product of a double-winding loop and a double-plaquette with the same clockwise orientation: 
Integration is performed over the link variable $V$ on the link which is common to the double-winding loop $W(D_2\times D_2)$ along the loop $D_2$ (the second term of eq.(\ref{ap-B-4})) and  the double-plaquette $\{{\rm tr}(V_{p}^{\dagger})\}^2$ adjacent to the common link $V$.
Here we have used the decomposition 
$W(D_2\times D_2):={\rm tr}(AV^{\dagger}AV^{\dagger})$ 
and 
${\rm tr}(V_{p}^{\dagger}):={\rm tr}(BV)$. 
}
\label{appendixB-fig2}
\end{center}
\end{figure}

Moreover, we proceed to perform the integration over the link variable for the product of a double-winding loop in $\tilde{W}_2$ and an adjacent double-plaquette $\{{\rm tr}(V_{p}^{\dagger})\}^2$.
As shown in Fig.~\ref{appendixB-fig2},  performing the integration of the link variable $V$ on the link which is common to the double-winding loop $W(D_2\times D_2)$ (the second term of eq.(\ref{ap-B-4})) and the double-plaquette $\{{\rm tr}(V_{p}^{\dagger})\}^2$ adjacent to the common link $V$ by using eq.(\ref{eq:BP3}) and eq.(\ref{A-sub2-3-3}), 
we obtain
\begin{align}
 \int dV \ W(D_2\times D_2) \cdot \left\{ {\rm tr}(V_{p}^{\dagger}) \right\}^2 
&=  \int dV \ {\rm tr}(AV^{\dagger}AV^{\dagger})\cdot \left\{{\rm tr}(BV)\right\}^2 
\nonumber \\
&= 
-\frac{2}{N(N^2-1)}\left\{  {\rm tr}(AB) \right\}^2  
+ \frac{2}{N^2-1}{\rm tr} (ABAB) 
\\
&:= 
-\frac{2}{N(N^2-1)}W(D_3)^2
+\frac{2}{N^2-1}W(D_3\times D_3) ,
\label{ap-B-6} 
\end{align}
where $D_3$ represents the loop as the boundary of a $3\times 1$ rectangle obtained by combining a $2\times 1$ rectangle and a plaquette adjacent to the common link $V$.
Then $W(D_3)$ and $W(D_3\times D_3)$  respectively stand for 
 the single-winding Wilson loop and double-winding Wilson loop along the loop $D_3$.
On the other hand, since the $V$ integral for the product of the first term of eq.(\ref{ap-B-4}), i.e., $W(D_2)^2$ and the double-plaquette variable adjacent to $V$, namely, 
$\int dV \ W(D_2)^2 \cdot \left\{ {\rm tr}(V_{p}^{\dagger}) \right\}^2$
 is the same type as eq.(\ref{ap-B-4}), we see that the result is again a linear combination of $W(D_3)^2$ and 
$W(D_3\times D_3)$.
Therefore, defining $\tilde{W}_3$ by the result of integration over the common link variable $V$ for the product of $\tilde{W}_2$ and the double-plaquette adjacent to the link $V$, namely,
$\tilde{W}_3:=\int dV \ \tilde{W}_2 \cdot \left\{ {\rm tr}(V_{p}^{\dagger}) \right\}^2$,
 we find $\tilde{W}_3$ is written as a linear combination of $W(D_3)^2$ and $W(D_3\times D_3)$.

From the above consideration, 
defining $\tilde{W}_n$ by the result of  connecting $n$ adjacent double-plaquettes one after another by integrating over the link variables inside the $S_2$ area, we can conclude that $\tilde{W}_n$ is written as
\begin{align}
\tilde{W}_n = \alpha_n W(D_n)^2 + \beta_n W(D_n\times D_n).
\label{ap-B-7}
\end{align}
This statement is proved by the mathematical induction. 
Indeed, by applying the same procedures as those given in eq.(\ref{ap-B-4}) and eq.(\ref{ap-B-6}) to eq.(\ref{ap-B-7}), we find the relationship 
\begin{align}
\tilde{W}_{n+1} 
:=& \int dV \ \tilde{W}_{n} \cdot \left\{ {\rm tr}(V_{p}^{\dagger}) \right\}^2
\nonumber\\  
 =& \left\{ \frac{2\alpha_n}{N^2-1} -\frac{2\beta_n}{N(N^2-1)} \right\} W(D_{n+1})^2 
+  \left\{  -\frac{2\alpha_n}{N(N^2-1)}+\frac{2\beta_n}{N^2-1} \right\} W(D_{n+1}\times D_{n+1})  
\nonumber\\  
 :=& \alpha_{n+1}W(D_{n+1})^2 +\beta_{n+1}W(D_{n+1}\times D_{n+1}).
\label{ap-B-8} 
\end{align}
Therefore, we have obtained the recurrence relation which holds for the coefficients $\alpha_n$ and $\beta_n$ for $n\geq 1$:
\begin{align}
\left( 
\begin{array}{c}
\alpha_{n+1} \\ -\beta_{n+1}
\end{array}
\right)
=\frac{2}{N^2-1}
\left( 
\begin{array}{cc}
1 & 1/N \\ 
1/N & 1
\end{array}
\right)
\left( 
\begin{array}{c}
\alpha_{n} \\ -\beta_{n}
\end{array}
\right).
\label{ap-B-9} 
\end{align}
Solving this  recurrence relation with the initial condition eq.(\ref{ap-B-5}),  we obtain the explicit form for the coefficients $\alpha_n$ and $\beta_n$ :
\begin{align}
\left( 
\begin{array}{c}
\alpha_n \\ -\beta_n
\end{array}
\right)
=2^{n-2}
\left( 
\begin{array}{c}
\left[  \frac{1}{N(N-1)}\right]^{n-1} + \left[  \frac{1}{N(N+1)}\right]^{n-1} \\ 
\left[  \frac{1}{N(N-1)}\right]^{n-1} - \left[  \frac{1}{N(N+1)}\right]^{n-1} 
\end{array}
\right).
\label{ap-B-10} 
\end{align}
Because the expansion coefficient $\frac{1}{2!}\left(\frac{1}{g^2}\right)^2$ is applied to each double-plaquette
in eq.(\ref{ap-B-3}),  a factor of $\frac{1}{2!^n}\left(\frac{1}{g^2}\right)^{2n}$ is applied to $n$ double-plaquettes.

\begin{figure}[tbp]
\begin{center}
\includegraphics[height=2.0cm]{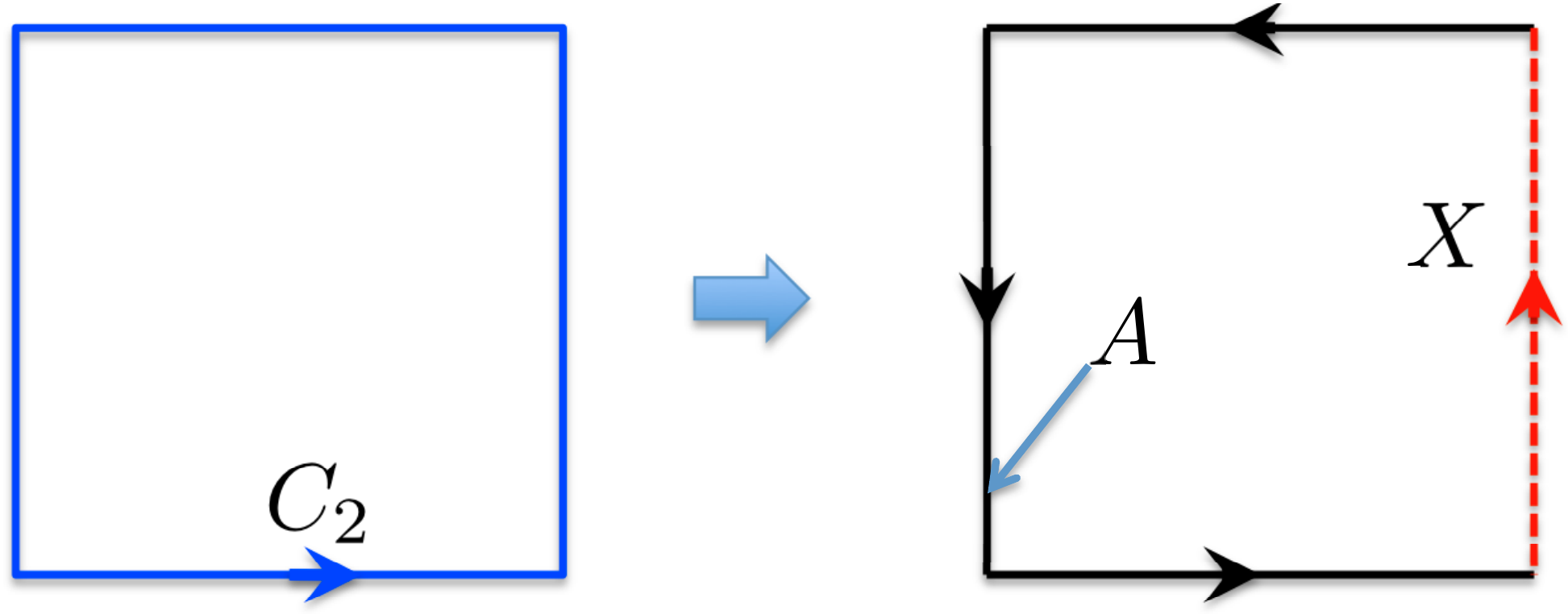}
\caption{ 
The path-ordered product of the link variables along the loop $C_2$ is decomposed into $A$ and $X$ to express $W(C_2\times C_2):={\rm tr}(AXAX)$.
}
\label{appendixB-fig3}
\end{center}
\end{figure}

Finally, we perform the integration over the remaining link variables on the loop $C_2$ as the boundary of the $S_2$ area.
As shown in Fig.\ref{appendixB-fig3}, we express  $W(C_2\times C_2)$  as $W(C_2\times C_2):={\rm tr}(AXAX)$.
To summarize the above arguments, from eq.(\ref{ap-B-2}), eq.(\ref{ap-B-3}), eq.(\ref{ap-B-7}) and eq.(\ref{ap-B-10}) etc., 
$\langle W(C_1\times C_2 ) \rangle_{q_N}$ is written by
\begin{align}
 \langle W(C_1\times C_2 ) \rangle_{q_N}   = 
  \left(  \frac{1}{g^2N} \right)^{S_1-S_2}\cdot \frac{1}{4}\left( \frac{1}{g^2}\right)^{2S_2}
  \int dAdX  \ {\rm tr}(AXAX)
  \left\{  x [ {\rm tr}(A^{\dagger}X^{\dagger})]^2 -y {\rm tr}(A^{\dagger}X^{\dagger}A^{\dagger}X^{\dagger})\right\},
\label{ap-B-11}
\end{align}
where  
\begin{align}
& x := \left[  \frac{1}{N(N-1)}\right]^{n-1} + \left[  \frac{1}{N(N+1)}\right]^{n-1} , 
\quad
y := \left[  \frac{1}{N(N-1)}\right]^{n-1} - \left[  \frac{1}{N(N+1)}\right]^{n-1} . 
\end{align}
Using eq.(\ref{eq:BP3}) and eq.(\ref{eq:BP4}) to perform $X$ integration, we finally obtain 
\begin{align}
 \langle W(C_1\times C_2 ) \rangle_{q_N}   =& 
 q_N \left(  \frac{1}{g^2N} \right)^{S_1+S_2},
\nonumber\\
 q_N =& -\frac{N^{2S_2}}{2} 
\left\{ 
\left[ \frac{1}{N(N-1)}\right]^{S_2-1} - \left[   \frac{1}{N(N+1)} \right]^{S_2-1}
\right\},  \ (S_2\geq 1),
\label{ap-B-13}
\end{align}
where $S_2\geq 1$ comes from the condition $n \geq 1$ in eq.(\ref{ap-B-9}).
It is easily checked that $\langle W(C_1\times C_2 ) \rangle_{q_N} =0$ when $S_2 =1$ by using 
explicit group integration.

\section{Explicit calculation of the coefficient $p_3$}
\crefname{equation}{}{}
\crefrangelabelformat{equation}{(#3#1#4--#5#2#6)}
\crefname{figure}{Fig.}{Figs.}
\setcounter{figure}{0}

In this section, we show explicitly how eq.(\ref{sce-su3-2}) is obtained. 
From eq.(\ref{chap3-3}) and eq.(\ref{chap3-5-2}), a contribution to a coplanar double-winding Wilson loop average 
$\langle W(C_1\times C_2 ) \rangle$ from the top panel of Fig.\ref{Dw-fig7} is expressed as 
\begin{align}
 \langle W(C_1\times C_2 ) \rangle_{q_3}   = \int\prod_{\ell \in S_1}dU_{\ell} \ W(C_1\times C_2 ) \cdot
 \prod_{p_j\in (S_1-S_2)} \left[  \frac{1}{g^2} {\rm tr} (U^{\dagger}_{p_j}) \right] \cdot
 \prod_{p_k\in S_2} \left[  \frac{1}{g^2} {\rm tr} (U_{p_k}) \right],
\label{ap-C-1}
\end{align}
where $U^{\dagger}_{p_j}$ and $U_{p_k}$ stand  respectively for plaquette variables on the $(S_1-S_2)$ and $S_2$ areas. 
Here note that $U_{p}^\dagger$ and $U_{p}$ respectively represent the plaquette variables for the plaquette ${p}$ with clockwise and counterclockwise orientations. 
In this section, we focus on the $N=3$ case. 

First, the integration with respect to the link variables $\{ U_\ell \}$ on the $(S_1-S_2)$ area can be performed with the same 
technique of the strong coupling expansion as that for the fundamental Wilson loop to obtain 
\begin{align}
 \langle W(C_1\times C_2 ) \rangle_{p_3}   = 
 \left(  \frac{1}{g^2N} \right)^{S_1-S_2}
  \langle W(C_2\times C_2 ) \rangle_{p_3} ,
\label{ap-C-2}
\end{align}
where we have defined 
\begin{align}
 \langle W(C_2\times C_2 ) \rangle_{p_3} := \int\prod_{\ell \in S_2}dU_{\ell} \ W(C_2\times C_2 ) \cdot
 \prod_{p_k\in S_2}  \left[  \frac{1}{g^2} {\rm tr} (U_{p_k}) \right].
\label{ap-C-3}
\end{align}

\begin{figure}[tbp]
\begin{center}
\includegraphics[height=1.0cm]{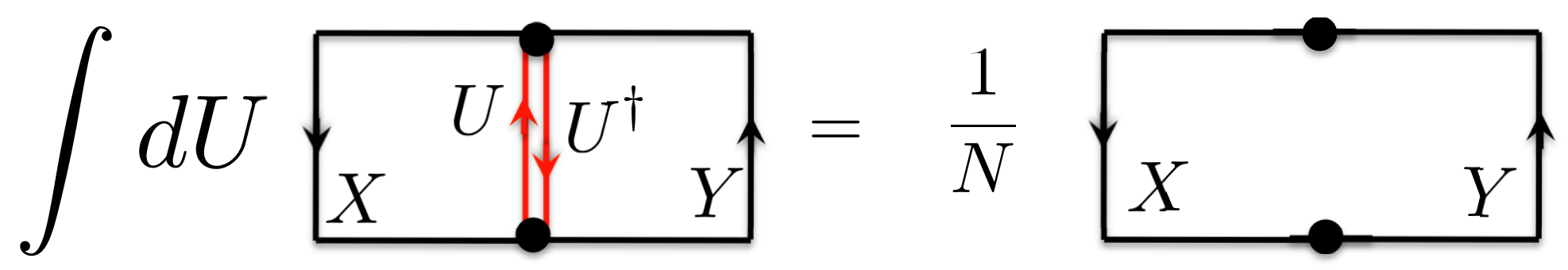}
\caption{ 
Diagrammatic representation of the integration rule eq.(\ref{ap-C-4}) for the product of two plaquettes with the same counterclockwise orientation:
Integration is performed over the link variable $U$ on the link which is common to two plaquettes with the same  counterclockwise orientation.
The plaquette variables for the plaquette $p_1$ and $p_2$ to the left and right of $U$ is respectively represented by 
${\rm tr}(U_{p_1}):={\rm tr}(XU)$ and ${\rm tr}(U_{p_2}):={\rm tr}(U^{\dagger}Y)$.
Here $X$ and $Y$ represent the products of the link variables along staple-shaped paths with the same orientations.
}
\label{appendixC-fig1}
\end{center}
\end{figure}

Next, we perform the integration in eq.(\ref{ap-C-3}) over link variables $\{ U_\ell \}$   inside of the $S_2$ area, which excludes the links on the loop $C_2$ as the boundary of the $S_2$ area.
As shown in Fig.\ref{appendixC-fig1}, performing the integration over the link variable $U$  using eq.(\ref{eq:BP1}) for two plaquettes that have a common link $U$, we obtain
\begin{align}
\int dU \ {\rm tr}(XU)\cdot {\rm tr}(U^{\dagger}Y)
= \frac{1}{N}  {\rm tr}(XY).
\label{ap-C-4} 
\end{align}

From this observation, we conclude that one factor of $1/N$ appears if two plaquettes are connected after common links are integrated.
When $S_2$ plaquettes are connected one after another by using eq.(\ref{ap-C-4}), a factor of $(1/N)^{S_2-1}$ is applied, and after that only the path ordered product of the link variables on the loop $C_2$ as the boundary of $S_2$ is left unintegrated. 
Therefore, eq.(\ref{ap-C-2}) becomes
\begin{align}
 \langle W(C_1\times C_2 ) \rangle_{p_3}   = 
  \left(  \frac{1}{g^2N} \right)^{S_1-S_2}\cdot N \left(  \frac{1}{g^2N} \right)^{S_2}
  \int_{[U]\in C_2} d[U]  \ W(C_2\times C_2)\cdot W(C_2),
\label{ap-C-5}
\end{align}
where the integral is only for the link variable on the loop $C_2$.

As shown in Fig.\ref{appendixB-fig3},  by using the decomposition $W(C_2):={\rm tr}(AX)$ and $W(C_2\times C_2):={\rm tr}(AXAX)$, and by repeatedly using eq.(\ref{A-sub2-5-4}), we obtain 
\begin{align}
 \int_{[U]\in C_2} d[U] \ W(C_2\times C_2)\cdot W(C_2) 
&= \int dAdX  \ {\rm tr}(AXAX){\rm tr}(AX) 
\nonumber \\ & 
= \int dAdX  \ {\rm tr}(XAXA){\rm tr}(XA) 
\nonumber \\
&= \int dAdX  \ (X)_{ab}(A)_{bc}(X)_{cd}(A)_{da}\cdot (X)_{pq}(A)_{qp} 
\nonumber \\ & 
= \frac{1}{N!}\epsilon_{acp}\epsilon_{bdq}\int dA  \ (A)_{bc}(A)_{da}(A)_{qp} 
\nonumber \\   
&=  \frac{1}{N!}\epsilon_{acp}\epsilon_{bdq} \cdot \frac{1}{N!}\epsilon_{bdq}\epsilon_{cap}  
= -1,
\label{ap-C-6}
\end{align}
where we have used the cyclicity of the trace  in the second equality.
Note that this result is meaningful only when $N=3$, because we have used eq.(\ref{A-sub2-5-4}) in the 
above calculation, eq.(\ref{A-sub2-5-5}) holds for $M\ne 0$ $({\rm mod} 3)$.
For $N=3$, thus, we obtain
\begin{align}
 \langle W(C_1\times C_2 ) \rangle_{p_3}   = 
  -3 \left(  \frac{1}{3g^2} \right)^{S_1},
 \label{ap-C-7}
\end{align}
which indeed yields $p_3=-3$.

\end{widetext}




\end{document}